\documentclass[sigconf]{acmart}

\usepackage{booktabs}
\usepackage{diagbox}
\usepackage{bm}
\usepackage{amsmath}

\usepackage{amssymb,amsfonts}
\usepackage[linesnumbered, ruled, vlined]{algorithm2e}
\usepackage{caption}
\usepackage{bigstrut}
\usepackage{subfig}
\usepackage{enumitem}
\usepackage{float}
\usepackage{balance}
\usepackage{multirow}
\usepackage{ulem}
\usepackage{tabularx}
\usepackage{pdfpages}
\usepackage{graphicx} 
\theoremstyle{definition}

\usepackage{threeparttable}
\usepackage{makecell}
\usepackage{cleveref}
\usepackage{array}
\usepackage{subcaption}
\usepackage{subfig}

\newcounter{anacounter}
\newcommand{\analysis}[0]{%
	\stepcounter{anacounter}%
	\textit{Analysis \theanacounter. }%
}

\newcounter{obscounter}
\newcommand{\observation}[1]{%
	\stepcounter{obscounter}%
	\textbf{\underline{Observation \theobscounter.} #1}%
}

\AtBeginDocument{%
  }

\setcopyright{acmlicensed}
\copyrightyear{2018}
\acmYear{2018}
\acmDOI{XXXXXXX.XXXXXXX}
\acmConference[Conference acronym 'XX]{Make sure to enter the correct
  conference title from your rights confirmation email}{June 03--05,
  2018}{Woodstock, NY}
\acmISBN{978-1-4503-XXXX-X/2018/06}

\setcopyright{none}
\renewcommand\footnotetextcopyrightpermission[1]{}
\begin{document}

\title{GPU-Accelerated Algorithms for Graph Vector Search: Taxonomy, Empirical Study, and Research Directions}


\author{
	Yaowen Liu{$^{1, *}$}, Xuejia Chen{$^{1, *}$}, Anxin Tian{$^2$},  Haoyang Li{$^1$}, Qinbin Li{$^3$}, Xin Zhang{$^2$} \\ Alexander Zhou{$^1$}, Chen Jason Zhang{$^1$}, Qing Li{$^1$, Lei Chen{$^2$}}
}
\thanks{$^*$ denotes equal contribution.}
\affiliation{
	\institution{$^1$Hong Kong Polytechnic University, Hong Kong SAR}
	\city{$^2$The Hong Kong University of Science and Technology, Hong Kong SAR \\ $^3$ Huazhong University of Science and Technology, China}
	\country{}
}
\renewcommand{\shortauthors}{Liu et al.}

\begin{abstract}
Approximate Nearest Neighbor Search (ANNS) underpins many large-scale data mining and machine learning applications, with efficient retrieval increasingly hinging on GPU acceleration as dataset sizes grow.
Although graph-based approaches represent the state of the art in approximate nearest neighbor search, there is a lack of systematic understanding regarding their optimization for modern GPU architectures and their end-to-end effectiveness in practical scenarios
In this work, we present a comprehensive survey and experimental study of GPU-accelerated graph-based vector search algorithms. We establish a detailed taxonomy of GPU optimization strategies and clarify the mapping between algorithmic tasks and hardware execution units within GPUs. 
Through a thorough evaluation of six leading algorithms on eight large-scale benchmark datasets, we assess both graph index construction and query search performance.
Our analysis reveals that distance computation remains the primary computational bottleneck, while data transfer between the host CPU and GPU emerges as the dominant factor influencing real-world latency at large scale. We also highlight key trade-offs in scalability and memory usage across different system designs. 
Our findings offer clear guidelines for designing scalable and robust GPU-powered approximate nearest neighbor search systems, and provide a comprehensive benchmark for the knowledge discovery and data mining community.

%

\end{abstract}

%


\received{20 February 2007}
\received[revised]{12 March 2009}
\received[accepted]{5 June 2009}

\maketitle

\section{Introduction}\label{sec:intro}

K-Nearest Neighbor Search (KNNS) is a core primitive in data mining, information retrieval, and vector search. 
Formally, given a dataset $\mathbf{X} = \{\bm{\mathit{x}}_1, \bm{\mathit{x}}_2, \ldots, \bm{\mathit{x}}_n\}$ consisting of $n$ data instances, where $\bm{\mathit{x}}_i \in \mathbb{R}^d$, and a query point $ \bm{\mathit{q}} \in \mathbb{R}^d$, the task is to retrieve the set of $k$ points in $\mathbf{X}$ nearest to $ \bm{\mathit{q}}$ according to a given distance metric. 
The standard exact solution evaluates the distance from $ \bm{\mathit{q}}$ to each of the $n$ points, resulting in a per-query time complexity of $O(nd)$. 
Although $O(nd)$ may appear efficient for small-scale data, in many real-world scenarios such as recommendation systems and retrieval-augmented generation, both $n$ and $d$ can be extremely large (often $n \ge 10^6$ and $d \ge 100$).
Therefore,  there is a critical requirement for real-time response to user queries. 
Under these conditions, even linear-time algorithms become impractical, as they cannot deliver required low latency or scale to to the volume and dimensionality of modern dataset sizes~\cite{DEEP, IVF} .




To address these scalability challenges, Approximate Nearest Neighbor Search (ANNS) relaxes the strict condition of exactness. By selecting approximate nearest neighbors, it trades a negligible loss in accuracy for significant gains in retrieval efficiency. 
The pursuit of effective ANNS has driven methodological evolution, beginning with tree-based~\cite{bentley1975kdtree, friedman1977findingbestmatch, arya1998bbdtree, silpa2008optimisedkdtree, muja2014scalable, chen2018kmeansforest, beis1997bestbinfirst, ram2019revisitingkdtree} structures that partition the search space into hierarchical regions to achieve logarithmic search complexity.
However, these methods suffer from severe performance degradation in high-dimensional spaces due to the inevitable overlapping of regions. To mitigate this, hash-based~\cite{zheng2020pm,indyk1998lsh, andoni2008optimalhashing, andoni2015optimallsh, weiss2008spectralhashing, liu2012supervisedhashing, cao2017hashnet, cao2018binaryhashing, shen2015superviseddiscretehashing, luo2023surveyonhash} methods were developed to project high-dimensional points into compact binary codes for sub-linear retrieval, yet they often struggle with precision loss and high sensitivity to hash function parameters. Alternatively, quantization-based~\cite{herve2011pq, ge2014optimizedpq, babenko2014additivequantization, zhang2014composite, guo2020accelerating, klein2019endtoend,su2020survey} methods compress vectors into a finite set of codewords to minimize storage and accelerate distance calculations, though they frequently encounter substantial computational overhead during the codebook training.

To maintain high recall while achieving superior search efficiency, graph-baseds~\cite{oguri2025graphreordering, xu2023spfresh, 2019diskann, chen2021spann, zhao2023lshapg, zhong2025enhancegraph, gong2025aannsinhg, zhao2023towards} methods have been proposed and have become the state-of-the-art for ANNS. The core principle of these methods lies in constructing a proximity graph where a greedy search algorithm navigates through nodes, iteratively exploring neighbors to converge on the k-nearest points of query.
However, early graph algorithms~\cite{2018HNSW, chen2018sptag, 2017NSG, hajebi2011knn, harwood2016fanng, fu2016efannaextremelyfast, malkov2014NSW} were primarily optimized for CPU. 
While GPUs can process multiple queries concurrently through multi-threading, the graph traversal within each individual query remains largely serial, as neighbor expansion, distance computation, and comparisons are performed sequentially.
To overcome this limitation, several GPU-accelerated schemes~\cite{guo2020accelerating, johnson2019billion} have emerged. Their fundamental approach is to leverage the massive parallelism of GPUs to perform concurrent distance computations and parallel node expansions, significantly boosting the throughput of large-scale vector retrieval.

Despite rapid progress in GPU-accelerated graph-based ANNS algorithms, several critical challenges continue to hinder rigorous evaluation and broader adoption in practical settings. In particular, prior works~\cite{wang2021comprehensive, Azizi2025survey} have primarily focused on CPU-based methods, while existing studies of GPU-based approaches~\cite{jiang2025gpusurvey, johnson2019billion} leave significant gaps in both methodological analysis and practical performance assessment. Specifically:

\begin{itemize}[leftmargin=8pt]
	\item \textbf{L1: Lack of a Structured Taxonomy.}  Existing research does not offer a unified taxonomy to categorize and summarize the wide range of GPU-accelerated techniques. Although GPUs provide massive parallelism, realizing their potential requires specialized optimizations that are neither systematically studied nor widely documented. Most prior work focuses on algorithmic results, with only limited discussion of core GPU strategies such as thread scheduling or memory management.
	
		\item \textbf{L2: Absence of a Unified Evaluation Framework.} There is no comprehensive framework designed to capture the full lifecycle and multifaceted performance aspects of GPU-based ANNS. Many studies concentrate primarily on search acceleration, with little attention paid to the associated overheads of graph construction or data movement that are critical in large-scale deployments.

		\item \textbf{L3: Incomplete End-to-End Evaluation.} Benchmarking practices remain fragmented, with performance commonly measured only by isolated kernel execution time. This approach ignores the considerable impact of memory allocation, PCIe data transfer, and host-device synchronization, leading to potentially misleading conclusions about real-world system performance.
	
\end{itemize}

{\color{black}
To bridge the critical gap between theoretical algorithmic design and practical hardware efficiency in GPU-accelerated graph-based ANNS, this work provides a systematic and multifaceted investigation into the current research landscape. Recognizing the lack of a unified understanding of how modern GPU architectures interact with high-dimensional search primitives, we first establish a comprehensive taxonomy of acceleration techniques and clarify the underlying mapping between algorithmic tasks and hardware execution units. We move beyond static analysis by conducting an extensive empirical study of six leading algorithms across eight large-scale benchmark datasets, carefully decomposing end-to-end performance to expose hidden bottlenecks. By synthesizing these theoretical insights with rigorous benchmarking, our study clarifies the complex trade-offs between computational throughput, data migration overhead, and memory scalability.


We summarize the contributions of this paper  as follows:

\begin{itemize}[leftmargin=15pt]
	\item We provide a structured taxonomy of GPU optimization strategies, offering a systematic guide for researchers to select effective acceleration techniques tailored for GPU-powered ANNS.
	
	\item We elucidate the mapping mechanisms between ANNS primitives and modern GPU hardware units, demystifying how architectural features can be leveraged for maximum efficiency.
	
	\item We evaluate the scalability and component runtime breakdown of state-of-the-art algorithms across datasets of varying sizes.
	
	\item We identify host-device data transfer latency as a critical performance bottleneck in large-scale retrieval environments.
\end{itemize}
}
 
%
%
%
%

\section{Preliminary}\label{sec:preliminaries}


This section establishes the fundamental concepts, notations, and problem formulations essential for understanding graph-based approximate nearest neighbor search (ANNS) algorithms and their implementation on GPU architectures.


\begin{definition}[{$k$-Nearest Neighbor Search.}]
Given a vector dataset  $\mathbf{X} = \{\bm{\mathit{x}}_1, \bm{\mathit{x}}_2, \ldots, \bm{\mathit{x}}_n\}$ 
 {where} $\bm{\mathit{x}}_i \in \mathbb{R}^d$, for a query point $\bm{\mathit{q}} \in \mathbb{R}^{d}$, the exact $k$-nearest neighbors $\mathcal{NN}_k(\bm{\mathit{q}})$ are defined as the set of $k$ points in $\mathbf{X}$ with the smallest distances to $q$:
\[
\mathcal{NN}_k(\bm{\mathit{q}}) = \{\bm{\mathit{x}}^{\bm{\mathit{q}}}_{1}, \bm{\mathit{x}}^{\bm{\mathit{q}}}_{2}, \ldots, \bm{\mathit{x}}^{\bm{\mathit{q}}}_{k}\} \subseteq \mathbf{X},
\]
where the indices $1, 2, \ldots, k$ satisfy
\[
\operatorname{dist}(\bm{\mathit{q}}, \bm{\mathit{x}}^{\bm{\mathit{q}}}_{j}) \leq \operatorname{dist}(\bm{\mathit{q}}, \bm{\mathit{x}}_\ell),\quad \forall\, j \in \{1, \ldots, k\},\, \forall\, \bm{\mathit{x}}_\ell \in \mathbf{X} \setminus \mathcal{NN}_k(\bm{\mathit{q}}),
\]
and $\operatorname{dist}(\cdot, \cdot)$ denotes the distance metric (e.g., Euclidean distance).
\end{definition}

 \begin{figure}[t]
	\centering
	\includegraphics[width=1.0\linewidth]{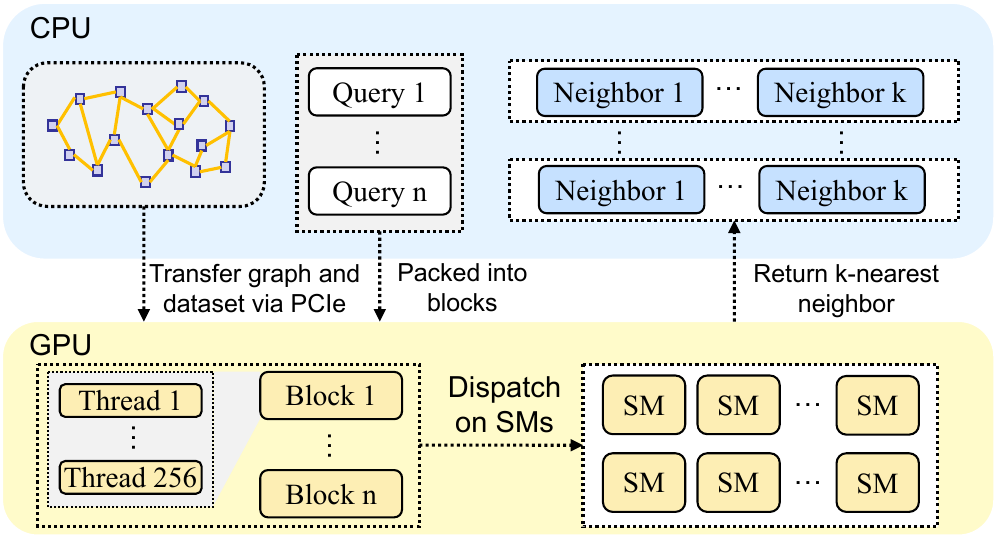}
	\caption{Workflow of GPU-Accelerated ANNS.}
	\label{fig:anns_on_gpu}
\end{figure}

\begin{definition}[Approximate \(k\)-Nearest Neighbor Search]
	The approximate version relaxes the requirement of finding the exact \(k\) nearest neighbors and instead seeks a set \(\mathcal{A}_k(\bm{\mathit{q}}) \subseteq \mathbf{X}\) with \(|\mathcal{A}_k(\bm{\mathit{q}})| = k\), where the selected elements are sufficiently close to the true neighbors. The effectiveness of approximate nearest neighbor search (ANNS) is commonly evaluated using recall, which quantifies the fraction of the true \(k\) nearest neighbors that are retrieved as
	$
	\mathrm{Recall@}k = \frac{|\mathcal{A}_k(\bm{\mathit{q}}) \cap \mathcal{NN}_k(\bm{\mathit{q}})|}{k},
	$
	where \(\mathcal{NN}_k(\bm{\mathit{q}})\) denotes the true \(k\) nearest neighbors of query \(\bm{\mathit{q}}\).
\end{definition}

{\color{black}
\noindent\textbf{GPU Architecture.}
As shown in Figure~\ref{fig:sm-arch} in the Appendix~\ref{ssec:gpu-arch}, a Graphics Processing Unit (GPU) is a specialized processor designed for large-scale parallel computations.
In the context of ANNS, the GPU's extensive parallelism and high memory bandwidth are crucial for accelerating computationally intensive operations such as high-dimensional distance calculations and large-scale  traversals. 
The GPU hardware hierarchy comprises several key components relevant to scalable ANNS, including the PCIe bus, streaming multiprocessors (SMs), and threads, detailed as follows:

\begin{itemize}[leftmargin=*]
	\item \textbf{PCIe:} The high-speed bus connecting the CPU (Host) and the GPU (Device). In ANNS, PCIe is responsible for migrating massive vector datasets and query batches to the GPU Global Memory. Minimizing PCIe overhead via asynchronous transfers is critical for hiding latency.

	\item \textbf{Streaming Multiprocessor (SM):} The fundamental hardware unit responsible for execution. A GPU contains multiple SMs, each equipped with CUDA cores, Tensor cores, and on-chip resources like registers and Shared Memory.
%
%

	\item \textbf{Thread:} The finest grain of the execution hierarchy. Each thread executes a sequence of instructions on a CUDA core. In a parallel ANNS implementation, individual threads handle specific vector dimensions or cooperatively manage candidate priority queues.
\end{itemize}
}

\noindent\textbf{ANNS on GPU.}
ANNS algorithms exploit GPU parallelism through distinct strategies in the index construction and search phases. During graph construction, data points are distributed across thread blocks, enabling parallel identification and establishment of neighbor connections, which greatly accelerates indexing compared to sequential methods. In the search phase (see Figure~\ref{fig:anns_on_gpu}), queries are processed in batches, with each assigned to a dedicated thread block for distance calculation and graph traversal. These blocks are dynamically scheduled across Streaming Multiprocessors (SMs), allowing concurrent execution of all queries in the batch and significantly improving system throughput.

\section{Component Analysis}\label{sec:components}

\textcolor{black}{This section categorizes and analyzes the build phase, and provides a detailed component-level and technique-level analysis of the search phase. The build phase is categorized into batch insertion, refinement, and divide-and-conquer approaches, with an analysis of their parallel acceleration strategies on GPU architectures. The search phase is decomposed into candidate locating, bulk distance computation, and data structure maintenance, with an in-depth examination of the parallel acceleration mechanisms and techniques employed by different algorithms.}

\begin{figure*}[htbp]
	\centering
	\includegraphics[width=0.9\textwidth]{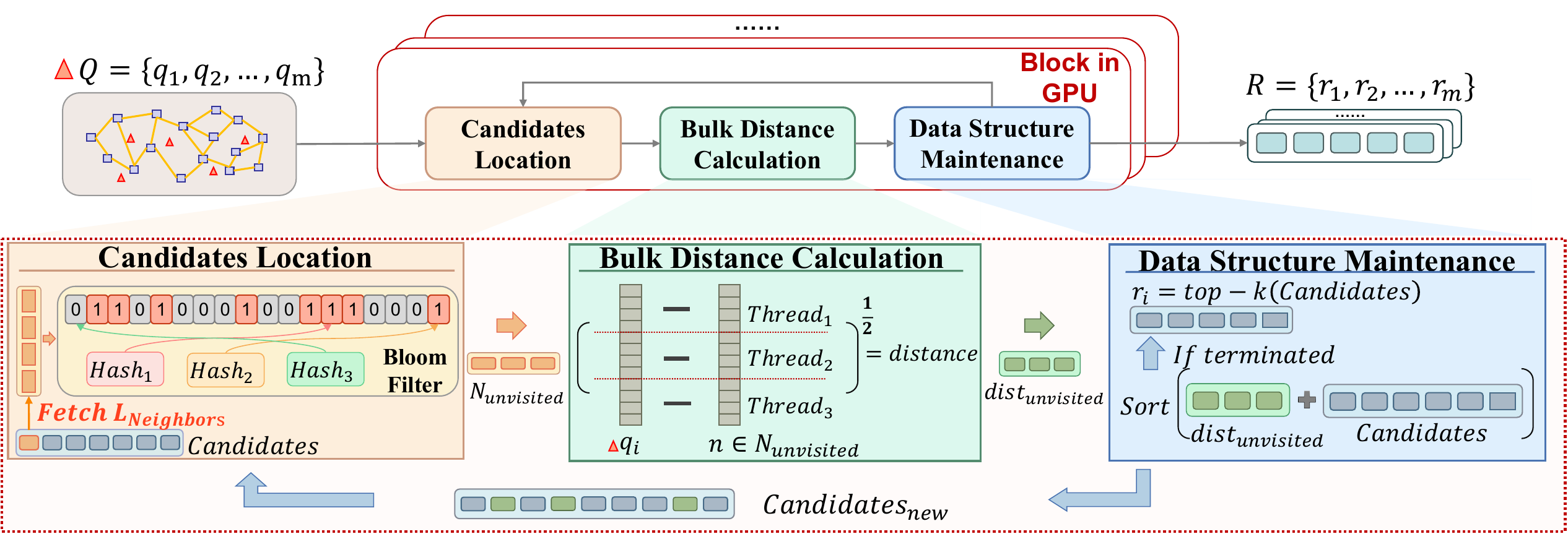}
	\caption{The pipeline of graph-based ANNS algorithms during the search phase.}
	\label{fig:component_for_search}
\end{figure*}


\subsection{Graph Construction Phase} \label{ssec:graph-construction-phase}


Due to architectural diversity, a universal construction framework remains elusive. In practice, the three primary methodologies, namely Concurrent Iterative Refinement (Type 1), Batch Increment (Type 2), and Parallel Divide and Conquer (Type 3), can be synergistically combined for superior performance. A detailed introduction to these types is provided in Appendix~\ref{ssec:graph-type}.

\noindent{\textbf{\underline{Type 1: Concurrent Iterative Refinement.}}}
This category follows an iterative graph refinement paradigm, where an initial approximate neighborhood graph is progressively improved through repeated neighbor exploration, re-ranking, and edge pruning. Typically, an initial KNNG is first generated using NN-descent~\cite{nn-descent} and then transferred to the GPU for the refinement phase. During this step, each vertex is assigned to an individual GPU block, which independently and in parallel identifies its N-hop neighbors and performs iterative refinement. After several iterations, the resulting high-quality approximate KNNG is transferred back to the CPU and stored on disk for subsequent search operations.
Compared to traditional CPU-based methods like NSG~\cite{2017NSG}, GPU-oriented refinement offers two major advantages. First, distance computations can be fully batched, enabling efficient utilization of GPU SIMD units and memory bandwidth. Second, because the refinement of each vertex is independent, the process allows for large-scale parallel execution and updates on the GPU, substantially improving graph construction efficiency.

Specifically,  {CAGRA~\cite{2024CAGRA}} uses rank-based reordering to count $N$-hop neighbors, where neighbors are reordered according to their relative connectivity ranks rather than purely by Euclidean distance. This rank-based reordering significantly speeds up the construction process and lowers memory consumption, and extensive experiments show that it preserves comparable search accuracy to distance-based refinement. {Tagore~\cite{li2025scalable}} uses Two-phase GNN-Descent~\cite{wang2021GNND} to optimize traditional NN-descent~\cite{nn-descent}. In the first phase, it retains the original NN-Descent~\cite{nn-descent} sample to refine the graph. In the second phase, each node samples from the neighbor lists of its m nearest neighbors to further pruning its neighbors.

\noindent{\textbf{\underline{Type2: Batch Increment.}}}
This category is derived from traditional CPU-based incremental insertion methods, where batches of points are inserted into the graph in parallel on the GPU. Each point is assigned to an individual GPU block to independently explore and identify its neighboring vertices.After all points have been inserted, the resulting graph is transferred back to the CPU and persisted to disk. Compared with conventional CPU-based incremental construction, it fully exploits the massive parallelism of GPU architectures, enabling simultaneous neighbor exploration for a large number of inserted points. This reduces graph construction latency and improves throughput, especially for large-scale datasets. Moreover, by processing insertions in batches, memory access patterns become more regular and cache-friendly, which further enhances GPU utilization and computational efficiency.

Specifically, GANNS\cite{2022GANNS} further improves construction quality by first constructing a local KNNG within each batch of inserted points. This intra-batch graph captures neighborhood relationships among newly inserted vertices, ensuring that close neighbors within the same batch are not overlooked during parallel insertion. By integrating the batch-level KNN structure into the global graph, GANNS~\cite{2022GANNS} effectively preserves search accuracy while maintaining high construction throughput on the GPU.

\noindent{\textbf{\underline{Type3: Parallel Divide and Conquer.}}}
This category leverages a divide-and-conquer paradigm to decompose large datasets into multiple subspaces, where local KNNG are constructed concurrently on the GPU. The built subgraphs are subsequently merged to generate the final global graph. This hierarchical parallelism enables efficient utilization of GPU resources and substantially accelerates graph construction for massive datasets while maintaining robust graph connectivity.
Specifically, GGNN~\cite{2022GGNN} follows a hierarchical framework that first constructs local subgraphs in parallel and then merges them in a top-down fashion, preserving graph quality comparable to sequential methods while delivering significant acceleration on GPUs. The top-level graph further serves as an initial seed structure to accelerate the subsequent search process.

\subsection{Search Phase} \label{ssec:search-phase}
As shown in Figure~\ref{fig:component_for_search}, the search process in graph-based ANNS follows a systematic pipeline that transforms the sequential traversal into parallelizable operations optimized for GPU architectures. Typically, we assign a warp (32 GPU threads) to process a single query point, and each warp executes independently without interfering with others. This pipeline decomposes the search into three fundamental stages that collectively enable efficient nearest neighbor discovery while leveraging GPU parallelism.
Algorithm~\ref{alg:greedy-search} in Appendix~\ref{ssec:alg-of-graph-search} illustrates the standardized greedy search procedure adapted for GPU. The algorithm demonstrates how the three pipeline stages are orchestrated in a parallel execution model.


\noindent{\underline{\textbf{S1: Candidates Locating in parallel.}}}
On GPU architectures, a set of threads collaboratively processes one candidate vertex. The threads jointly extract the nearest unvisited candidates and fetch its neighboring vertices in parallel. Since neighbors are stored at distinct memory offsets, concurrent loads from global memory do not incur write conflicts, allowing the fetched neighbors to be efficiently staged in shared memory. Compared with traditional CPU-based graph search, where candidate extraction and neighbor expansion are typically performed sequentially or with limited core-level parallelism and suffer from irregular memory accesses, the GPU-based design exploits fine-grained warp-level parallelism and high memory bandwidth to significantly accelerate candidate locating and improve overall search throughput.

Early GPU-based graph ANNS method SONG~\cite{2020SONG}, employ a single thread to fetch the neighbors of a candidate vertex, while the remaining threads within the thread block remain idle during this stage. In contrast, subsequent GPU-based methods, including GGNN~\cite{2022GGNN} and GANNS~\cite{2022GANNS}, significantly optimize this process by allowing all threads within a thread block to cooperatively fetch different neighbors in parallel. This parallel neighbor fetching strategy substantially improves memory throughput and reduces latency, leading to more efficient candidate location on GPUs. A notable exception is BANG~\cite{2024BANG}, where the graph data are not fully resident on the GPU. Consequently, neighbor fetching is performed on the CPU, and the fetched neighbor information is asynchronously transferred to the GPU for subsequent processing, which enables BANG~\cite{2024BANG} to handle large-scale graphs that exceed GPU memory capacity.

While almost all the algorithms adopt the bloom filter, which deletes the neighbors already computed,  GANNS~\cite{2022GANNS} observes that distance computation can be efficiently parallelized on the GPU and achieves significant speedup through fine-grained thread-level parallelism. However, applying a Bloom filter to  involves random memory accesses, which are detrimental to GPU memory bandwidth. Therefore, it eliminates the filtering mechanism entirely. PathWeaver~\cite{kim2025pathweaver} uses Direction-Guided Selection to reduces the number of distance calculations while maintaining high accuracy. The query-to-node direction is computed and compared with the table to count matching sign bits for all neighbors. Neighbors with the most matches are selected as candidates.

\noindent{\underline{\textbf{S2: Bulk Distance Calculation.}}}
Bulk distance computation is parallelized by partitioning the calculation of Euclidean distances across multiple threads, where each thread is responsible for a subset of the dimensions of the query and data points, and the partial results computed by each thread are finally reduced to obtain the complete Euclidean distance. Compared with CPU implementations, this GPU-based design achieves the most significant acceleration in graph-based ANNS, fully exploiting thread-level parallelism and high memory bandwidth.

Most algorithms like SONG~\cite{2020SONG}, GGNN~\cite{2022GGNN} uses a warp(a set of 32 threads) to compute the distance between a single data point and the query point. Differently,  CAGRA~\cite{2024CAGRA} introduces the Warp Splitting technique, which divides a warp into smaller groups of 8 threads, with each group responsible for computing the distance for a different point. This design increases GPU occupancy and parallelism, improving overall throughput. In our Appendix~\ref{parameters}, we experimentally evaluate different group sizes, and the results show that warp splitting achieves higher QPS compared to the traditional warp-per-point approach at the same recall level.

\noindent{\underline{\textbf{S3: Data Structure Maintenance.}}} 
 This stage updates the candidate list based on the results from the previous step. Typically, threads within the same warp or thread block are assigned to write to different positions in the priority queue or candidate data structure, which avoids write conflicts and maximizes GPU parallel efficiency. The GPU design allows threads to concurrently update multiple candidate entries, resulting in substantial acceleration in this stage of graph-based ANNS compared to traditional CPU methods.
In SONG~\cite{2020SONG} and PilotANN~\cite{2025PilotANN}, a heap structure is used to maintain the candidate list. When a distance is computed, only the first thread pushes it into the heap. GGNN~\cite{2022GGNN} employs a circular queue, where each thread is responsible for moving one element in the array when inserting a new point, which greatly accelerates the data structure updates. In contrast, other algorithms like GANNS~\cite{2022GANNS}, CAGRA~\cite{2024CAGRA} adopt a lazy check strategy, which avoids pushing points to the candidate list immediately after distance calculation. Instead, they store the distances in an array \(d_{\text{unvisited}}\) and then use bitonic sort, a GPU-based sorting algorithm to sort \(d_{\text{unvisited}}\) and merge it into the candidate list after all neighbor distances have been computed. Bitonic sort fully leverages each thread's computational capability compared to inserting nodes one by one.

\vspace{-5pt}
\section{Experimental Evaluations}\label{sec:exp}

This section presents a comprehensive experimental evaluation of GPU-accelerated graph-based ANNS algorithms. We first detail the experimental setup and metrics in Section~\ref{ssec:exp-setting}. Next, Section~\ref{ssec:construction-evaluation} assesses index construction efficiency and graph quality. Section~\ref{ssec:search-performance} forms the core analysis, decomposing search performance into end-to-end latency, kernel efficiency, and sub-component breakdowns. Finally, Section~\ref{ssec:scalability} examines scalability regarding data volume and query throughput.

\subsection{Experimental Settings}\label{ssec:exp-setting}

\noindent \textbf{Implementation setup.} All experiments are performed on a server running Ubuntu 22.04 LTS, equipped with an Intel Xeon Gold 6348 CPU (2.60 GHz, 112 threads) and an NVIDIA A800 GPU with 80 GB HBM2e memory. All algorithms are compiled with GCC 11.4.0 under the -O3 optimization flag, and CUDA 12.4 is employed for GPU acceleration throughout our experiments. The system is implemented in C++17 and CUDA C++.
\begin{table}[h]
	\centering
	\caption{Statistics of datasets.}
	\setlength{\intextsep}{10pt}
	\setlength{\tabcolsep}{2pt}
	\renewcommand{\arraystretch}{1.2}
	\label{tab:statistics-of-datasets}
	\begin{tabular}{lccccc}
		\toprule
		\textbf{Dataset} & \textbf{Volume} & \textbf{\#Dim} & \textbf{Type} & \textbf{Size (GB)} &\textbf{Metric}\\
		\midrule
		NYTimes\cite{2008NYTimes}    & 290,000      & 256   & float  & 0.301  & cos \\
		DEEP1M~\cite{DEEP}           & 1,000,000    & 96    & float  & 0.358  & L2 \\
		SIFT1M~\cite{IVF}            & 1,000,000    & 128   & float  & 0.492  & L2 \\
		GIST~\cite{IVF}              & 1,000,000    & 960   & float  & 3.58   & L2 \\
		GLoVe200~\cite{2014glove}    & 1,183,514    & 256   & float  & 0.918  & cos \\
		MNIST8M~\cite{lecun1998mnist}& 8,090,000    & 784   & float  & 24     & L2 \\
		DEEP10M~\cite{DEEP}          & 10,000,000   & 96    & float  & 3.58   & L2 \\
		DEEP100M~\cite{DEEP}         & 100,000,000  & 96    & float  & 35.8   & L2 \\
		\bottomrule
	\end{tabular}
\end{table}

\noindent \textbf{Datasets.} We evaluated GPU-accelerated graph-based ANNS algorithms on 8 benchmark datasets covering different scales, dimensions, and data types to ensure comprehensive assessment. Detailed statistics of each dataset are shown in Table~\ref{tab:statistics-of-datasets}.

\begin{itemize}[left=0pt]
    \item \textbf{NYTimes~\cite{2008NYTimes}.} The NYTimes dataset is a large-scale collection of news articles from The New York Times, containing rich textual content, metadata, and categorical information.
    \item \textbf{DEEP~\cite{DEEP}.} The Deep100M dataset consists of 100 million image embeddings extracted from the final layer of GoogLeNet, PCA-compressed to 96 dimensions and l2-normalized. Its smaller subsets, DEEP1M and DEEP10M, are constructed by taking the first 1 million and 10 million entries, respectively.
    \item \textbf{SIFT~\cite{IVF}.} SIFT is a benchmark dataset of 128-dimensional SIFT (Scale-Invariant Feature Transform) descriptors used for evaluating approximate nearest neighbor search algorithms.
    \item \textbf{GIST~\cite{IVF}.} GIST is a dataset containing GIST descriptors, which are low-dimensional global image features that capture the spatial envelope of scenes
    \item \textbf{GloVe200~\cite{2014glove}.} This dataset contains 1,200,000 word vectors,each with 200 dimensions from GloVe model.
    \item \textbf{MNIST8M~\cite{lecun1998mnist}.} It contains nearly 8 million data points, each represented by a 784-dimonsional float vector.
\end{itemize}

\noindent \textbf{Algorithms.} We evaluated 6 open-source GPU-accelerated ANNS algorithms, as detailed in section~\ref{ssec:graph-anns-gpu}. Since FlashANNS~\cite{xiao2025flashanns} only provides a preview version, it was not included in the testing. For Tagore~\cite{li2025scalable} and GRNND~\cite{li2025grnnd}, they focus on improving graph construction and graph quality and do not provide an independent search algorithm. To prevent confounding in our comparisons, we exclude these methods from Table~\ref{tab:trans}. To highlight the GPU acceleration effects, we compared them with representative CPU methods HNSW~\cite{2018HNSW} and VAMANA~\cite{2019diskann}, where HNSW~\cite{2018HNSW} was implemented using the Faiss library and VAMANA\cite{2019diskann} was implemented using the official DiskANN~\cite{2019diskann} repository.

\noindent \textbf{Parameters.} For build phase, we enforce the same maximum degree for all algorithms although some algorithms have an unfixed degree by design. For the search phase, we choose to search the top-10 nearest neighbors. We set different iteration times to get the results of QPS and recall. 

\begin{table*}[t]
	\centering
	\caption{Index construction metrics on medium to large-scale datasets (Part 1). CT: Construction Time, AD: Average Degree, CC: Connected Components, PMF: Peak Memory Footprint in MB (CPU/GPU). The symbol `--' means that PilotANN cannot work on NYTimes and GLoVe200 due to cosine distance unsupported.}
	\setlength{\tabcolsep}{2pt}
	\renewcommand{\arraystretch}{1.2}
	\resizebox{0.95\textwidth}{!}{
	\begin{tabular}{c|cccc|cccc|cccc|cccc|cccc}
		\hline
		\multirow{2}{*}{Alg} 
		& \multicolumn{4}{c|}{NYTimes~\cite{2008NYTimes}}  
		& \multicolumn{4}{c|}{SIFT1M~\cite{IVF}} 
		& \multicolumn{4}{c|}{DEEP1M~\cite{DEEP}}
		& \multicolumn{4}{c|}{GLoVe200\cite{2014glove}}
		& \multicolumn{4}{c}{GIST\cite{IVF}}   \\
		\cline{2-21}
		& CT & AD & CC & PMF
		& CT & AD & CC & PMF 
		& CT & AD & CC & PMF 
		& CT & AD & CC & PMF 
		& CT & AD & CC & PMF \\
		\hline
		SONG~\cite{2020SONG}      &  1409s & 48  & 326  & 710/0  & 972s  & 23  & 105  & 1105/0  & 759s  &  23 & 328  & 983/0  & 12183s  & 69  & 320  & 3345/0 & 33539s  & 88  & 52  & 5843/0   \\
		
		GGNN~\cite{2022GGNN}     & 22s  & 64  & 1  &  405/801 &  \textbf{6s$^{*}$} & 32  & 1  & 738/1171  & \underline{9s}  & 32  & 38  & 614/1049  & 338s  & 96  & 1  & 1475/2201 & 869s  & 128  & 11  & 4293/8243    \\
		
		GANNS~\cite{2022GANNS}     &  \textbf{3s$^{*}$} & 48  & 196  & 392/2402  & \underline{12s}  & 24  & 287  & 723/5523  & \textbf{7s$^{*}$}  & 24  & 623  & 598/5101  & \underline{65s}  & 92  & 56  & 1641/8057 & \underline{145s}  & 88  &  62 & 4260/9861    \\
		
		CAGRA~\cite{2024CAGRA}      & \underline{4s}  & 64  & 1  & 1577/1449  & \underline{12s}  & 32  & 1  & 4134/2575  & 10s  & 32  &  1 & 2792/3499  & \textbf{18s$^{*}$}  & 96  & 1  & 5157/4523 & \textbf{18s$^{*}$}  & 128  & 1  & 7675/9803   \\
		
		Pilot-ANN\cite{2025PilotANN}   & -  & -  & -  & -  & 1392s  &  21 & 1  & 5423/0  & 1195s  & 22  & 1  & 4397/0  & -   & -    & -   & -  & 4969s  & 95  & 1  & 15563/0   \\
		
		HNSW~\cite{2018HNSW}    & 48s  & 58  & 1  & 728/0  & 22s  & 27  & 1 & 1274/0  & 16s  & 23  & 1  & 1022/0   & 5960s  & 96  & 1  &  3016/0  & 5960s  & 96  & 1  &  3016/0  \\
		
		VAMANA~\cite{2019diskann}   &  787s & 64  &  1 & 1706/0  & 208s  & 32  & 1  & 4149/0  & 347s  & 32  & 1  & 2789/0  & 1533s  & 96  & 1  & 5560/0  & 5071s  & 128  & 1  & 14754/0  \\
		\hline
	\end{tabular}
	}
	\label{tab:construction-eff-part1}
\end{table*}

\noindent \textbf{Evaluation metrics.} To comprehensively evaluate the performance of the algorithm, we measure various metrics during both the build (index construction) and search (query processing) phases.
For the build phase, metrics are as follows:
\begin{itemize}[left=0pt]
    \item \textbf{Construction Time (CT).} The time taken to build the index.
    \item \textbf{Peak Memory Footprint (PMF).}The peak memory usage are recorded during the index construction phase, which includes both CPU and GPU memory.
    \item \textbf{Average Degree (AD).} The average out degree of the graph.
    \item \textbf{Connected Components (CC).} The number of connected components in the graph.
    \item \textbf{Path Length (PL).} The average number of hops during search.
\end{itemize}
For the search phase, metrics are as follows:
{\color{black}
\begin{itemize}[left=0pt]
	\item \textbf{Queries Per Second (QPS).} It quantifies the search throughput, defined as the ratio of the total number of processed queries ($N_{\mathbf{Q}}$) to the total elapsed time (T).
	
	\item \textbf{Recall@K.} It evaluates the search accuracy by measuring the fraction of true \(k\) nearest neighbors successfully retrieved in the set \(\mathcal{A}_k(\bm{\mathit{q}})\):
	$\mathrm{Recall@}k = \frac{|\mathcal{A}_k(\bm{\mathit{q}}) \cap \mathcal{NN}_k(\bm{\mathit{q}})|}{k}.$
\end{itemize}
}

{\color{black}
\subsection{Graph Construction Evaluation}\label{ssec:construction-evaluation}

\subsubsection{Construction Efficiency Evaluation}
In the context of approximate nearest neighbor search, graph construction efficiency is a pivotal factor that determines not only the initial deployment speed but also the frequency with which the system can be updated to accommodate new data. We evaluate the construction efficiency of seven representative algorithms across eight diverse datasets ranging from medium to large scales. The evaluation focuses on four key metrics: Construction Time (CT), Average Degree (AD), Connected Components (CC), and Peak Memory Footprint (PMF) on both CPU and GPU. Note that BANG is omitted from the tables as it adopts the same construction mechanism as VAMANA~\cite{2019diskann}.

As shown in Table~\ref{tab:construction-eff-part1} and Table~\ref{tab:construction-eff-part2} in Appendix~\ref{ssec:supplementary-of-construction}, GPU-accelerated algorithms significantly outperform traditional CPU-based methods in terms of construction speed.
The experimental results unequivocally demonstrate that leveraging GPU massive parallelism is essential for handling large-scale and high-dimensional datasets. Algorithms like CAGRA~\cite{2024CAGRA} and GGNN~\cite{2022GGNN}, which fully utilize GPU resources for parallel distance calculations and graph updates, achieve speedups of multiple orders of magnitude compared to sequential CPU execution. However, this speed comes at the cost of higher GPU memory consumption, as evidenced by the OOM$^{*}$ failures on the largest datasets for some methods.

\noindent\textit{\textbf{Insight:}} 
Future optimization must prioritize hybrid CPU-GPU strategies to resolve the critical trade-off between construction speed and memory constraints. By orchestrating host memory for massive graph storage while retaining GPU parallelism for computation, next-generation systems can effectively mitigate OOM failures and ensure scalability for billion-scale datasets.

\begin{table*}[t]
	\centering
	\caption{Data Transferring Time (Trans.) and Search Time at 90\% Recall@10. The symbol `--' indicates that the algorithm failed to run on the dataset.}
	\label{tab:trans}
	\setlength{\tabcolsep}{1.5pt}
	\renewcommand{\arraystretch}{1.5}
	\resizebox{0.95\textwidth}{!}{
		\begin{tabular}{c|cc|cc|cc|cc|cc|cc|cc|cc}
			\hline
			\multirow{2}{*}{Algorithm}
			& \multicolumn{2}{c|}{NYTimes}
			& \multicolumn{2}{c|}{SIFT1M}
			& \multicolumn{2}{c|}{Deep1M}
			& \multicolumn{2}{c|}{GIST}
			& \multicolumn{2}{c|}{GloVe}
			& \multicolumn{2}{c|}{Deep10M}
			& \multicolumn{2}{c|}{MNIST}
			& \multicolumn{2}{c}{Deep100M} \\
			\cline{2-17}
			& Trans. & Search
			& Trans. & Search
			& Trans. & Search
			& Trans. & Search
			& Trans. & Search
			& Trans. & Search
			& Trans. & Search
			& Trans. & Search \\
			\hline
			SONG~\cite{2020SONG}
			& 42.00 & 178.32
			& 72.67 & 71.46
			& 59.45 & 71.14
			& 722.73 & 51.04
			& 263.01 & 368.04
			& 810.71 & 70.44
			& 3774.14 & 280.92
			& -- & -- \\
			\hline
			GGNN~\cite{2022GGNN}
			& 38.22 & 110.58
			& 61.67 & 21.32
			& 48.33 & 21.51
			& 384.18 & 25.46
			& \textbf{131.98$^{*}$} & 495.99
			& 642.03 & 39.82
			& 3161.45 & 267.20
			& 7518.18 & 104.91 \\
			\hline
			GANNS~\cite{2022GANNS}
			& 36.58 & 214.44
			& 114.21 & 24.00
			& 99.44 & 18.94
			& 1474.81 & \textbf{21.31$^{*}$}
			& 276.29 & 412.08
			& 1325.24 & 28.17
			& 8545.01 & 114.15
			& -- & -- \\
			\hline
			CAGRA~\cite{2024CAGRA}
			& \textbf{35.73$^{*}$} & \textbf{41.10$^{*}$}
			& 119.05 & \textbf{11.00$^{*}$}
			& 48.21 & \textbf{11.77$^{*}$}
			& 450.54 & 23.96
			& 180.98 & \textbf{131.29$^{*}$}
			& 612.83 & \textbf{14.92$^{*}$}
			& 3423.75 & \textbf{49.98$^{*}$}
			& 5311.59 & \textbf{18.98$^{*}$} \\
			\hline
			BANG~\cite{2024BANG}
			& 707.40 & 9501.29
			& \textbf{17.61$^{*}$} & 159.02
			& \textbf{19.85$^{*}$} & 255.81
			& \textbf{11.21$^{*}$} & 187.11
			& 256.73 & 5412.72
			& \textbf{17.11$^{*}$} & 384.36
			& \textbf{65.71$^{*}$} & 674.28
			& \textbf{38.97$^{*}$} & 530.13 \\
			\hline
			PilotANN~\cite{2025PilotANN}
			& -- & --
			& 79.12 & 185.28
			& 57.57 & 166.19
			& 181.36 & 2108.11
			& -- & --
			& 381.89 & 297.11
			& 1430.11 & 1146.83
			& -- & -- \\
			\hline
		\end{tabular}
	}
\end{table*}

\begin{figure}[t]
	\vspace{-5pt}
	\centering
	\includegraphics[width=0.95\linewidth]{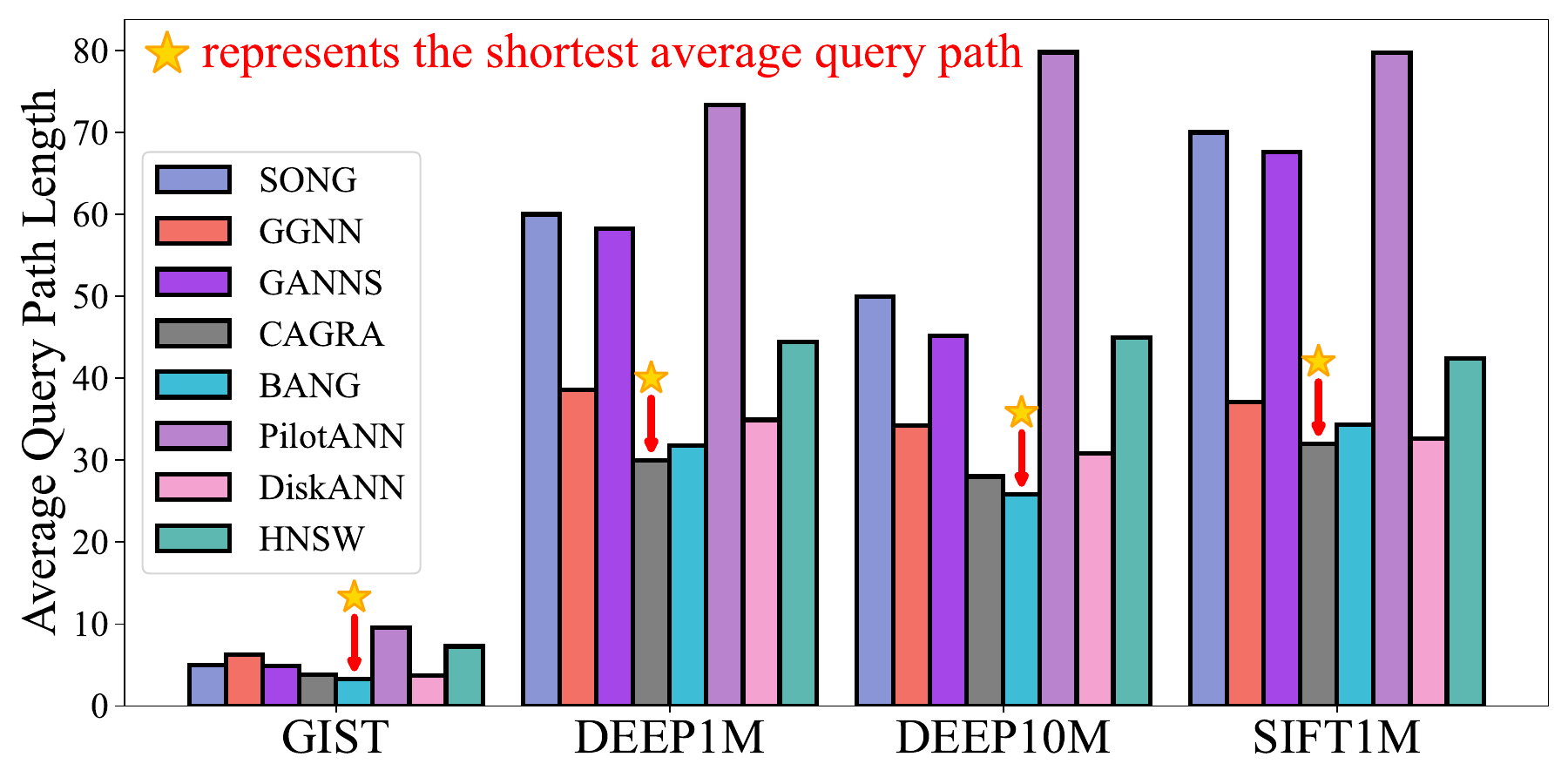}
	\caption{Average query path length across different algorithms at 90\% recall@10.}
	\label{fig:exp-graph-quality}
	\vspace{-10pt}
\end{figure}

\subsubsection{Graph Quality Evaluation}
Unlike previous surveys~\cite{wang2021comprehensive} that assess graph quality based on topological deviation from the exact KNNG, we adopt a search-oriented metric: the average query path length required to achieve a target accuracy. A high-quality graph should facilitate rapid convergence to the nearest neighbors through shorter greedy traversal paths. We evaluated this metric on four datasets with graph degrees fixed at 128 for GIST~\cite{IVF}, 64 for DEEP10M~\cite{DEEP}, and 32 for DEEP1M~\cite{DEEP} and SIFT1M~\cite{IVF},  respectively, measuring the path length when Recall@10 reaches 90\%.

As illustrated in Figure~\ref{fig:exp-graph-quality}, there is a significant variance in traversal efficiency among the evaluated algorithms. CAGRA~\cite{2024CAGRA} and BANG~\cite{2024BANG} consistently demonstrate superior graph quality, marked by the shortest query paths across all datasets. The results underscore that graph quality is defined by the "navigability" of the structure rather than simple connectivity. Algorithms that employ advanced pruning or edge selection strategies, such as CAGRA~\cite{2024CAGRA}, effectively create "highways" in the high-dimensional space, allowing greedy searches to converge with fewer hops.

\noindent\textit{\textbf{Insight:}} 
Future research should pivot towards enhancing graph "navigability" via hardware-aware pruning, ensuring that topological "highways" align with GPU memory access patterns to minimize traversal hops and random latency.
}

{\color{black}
\subsection{Search Performance} \label{ssec:search-performance}

\subsubsection{Search with Transfer}
We evaluate end-to-end latency by decomposing execution into Host-to-Device (HtoD) transfer and on-device search. The Device-to-Host (DtoH) phase is excluded, as transferring small top-k results incurs negligible and uniform latency. We benchmark six algorithms across diverse datasets ranging from 0.29M to 100M, measuring the total processing time for 10,000 queries at 90\% recall@10.

Table~\ref{tab:trans} reveals a severe bottleneck in data migration foTr pure GPU algorithms, including SONG~\cite{2020SONG}, GGNN~\cite{2022GGNN}, GANNS~\cite{2022GANNS}, CAGRA~\cite{2024CAGRA}. As dataset size increases, HtoD transfer time grows disproportionately, consuming approximately 40-80\% of total latency on DEEP1M~\cite{DEEP} and exceeding 97\% on DEEP100M~\cite{DEEP}, effectively marginalizing the benefits of fast kernel execution. Conversely, hybrid algorithms like BANG~\cite{2024BANG} and PilotANN~\cite{2025PilotANN} exhibit lower transfer times by leveraging CPU processing, but this trade-off results in higher search latencies due to host-side computational overhead. The dominance of HtoD transfer time underscores that PCIe bandwidth, rather than GPU compute capability, is the primary limiting factor for large-scale ANNS. Optimizing graph search kernels yields diminishing returns when data migration occupies over 90\% of the runtime.

\noindent\textit{\textbf{Insight:}} Future research must prioritize strategies to mitigate this "memory wall," such as asynchronous overlap of transfer and compute, aggressive vector compression during transport, or entirely resident-GPU indexing strategies to eliminate transfer costs.

\begin{figure}[t]
	\vspace{-10pt}
	\includegraphics[width=0.7\linewidth]{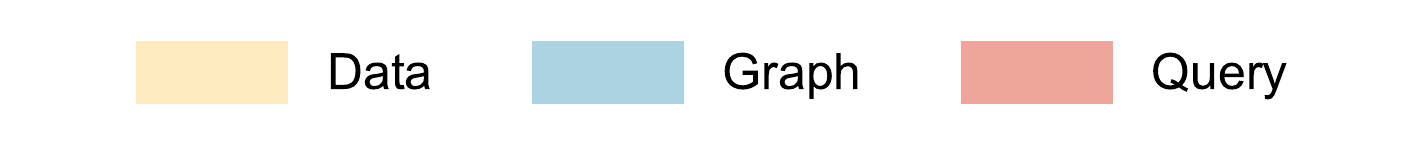}
	\vspace{-10pt}
\end{figure}

\begin{figure}[t]
	\vspace{-10pt}
	\centering
	\subfloat[CAGRA.]{
		\includegraphics[width=0.31\linewidth]{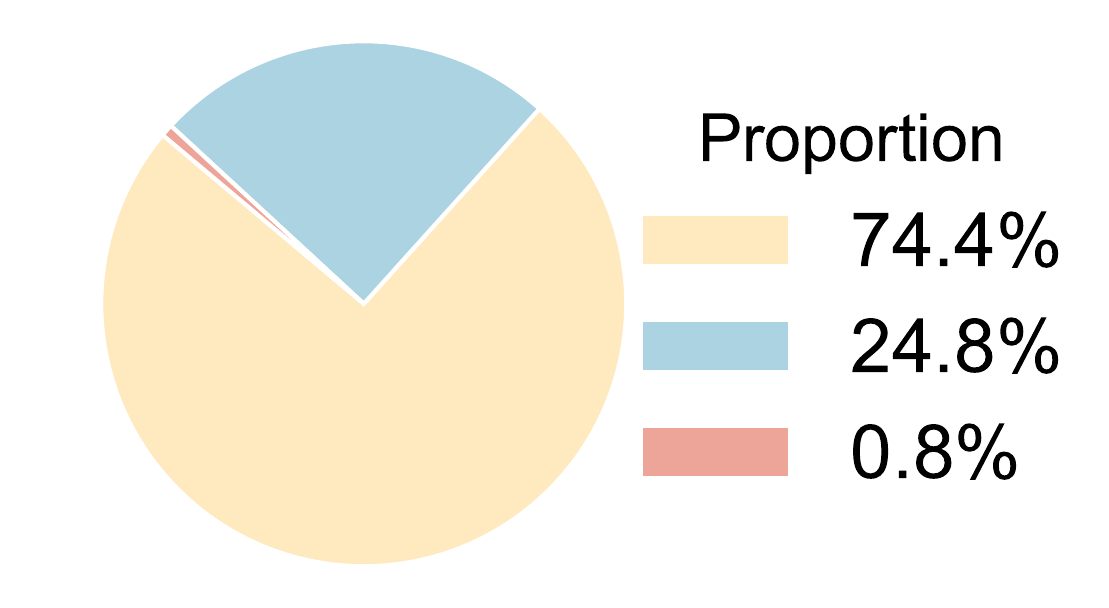}
	}
	\hfill
	\subfloat[BANG.]{
		\includegraphics[width=0.31\linewidth]{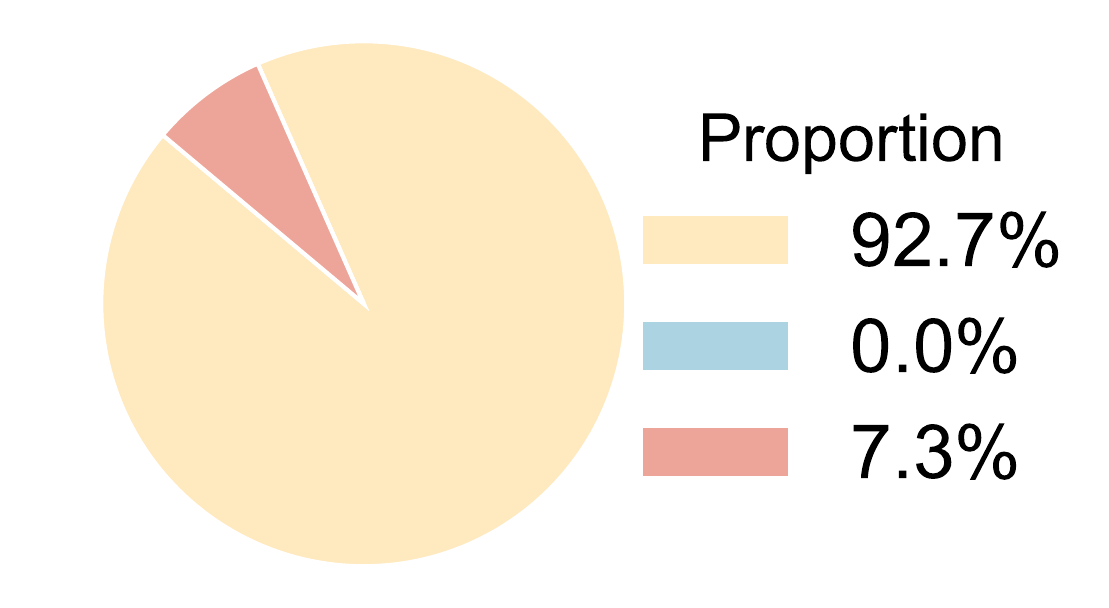}
	}
	\hfill
	\subfloat[PilotANN.]{
		\includegraphics[width=0.31\linewidth]{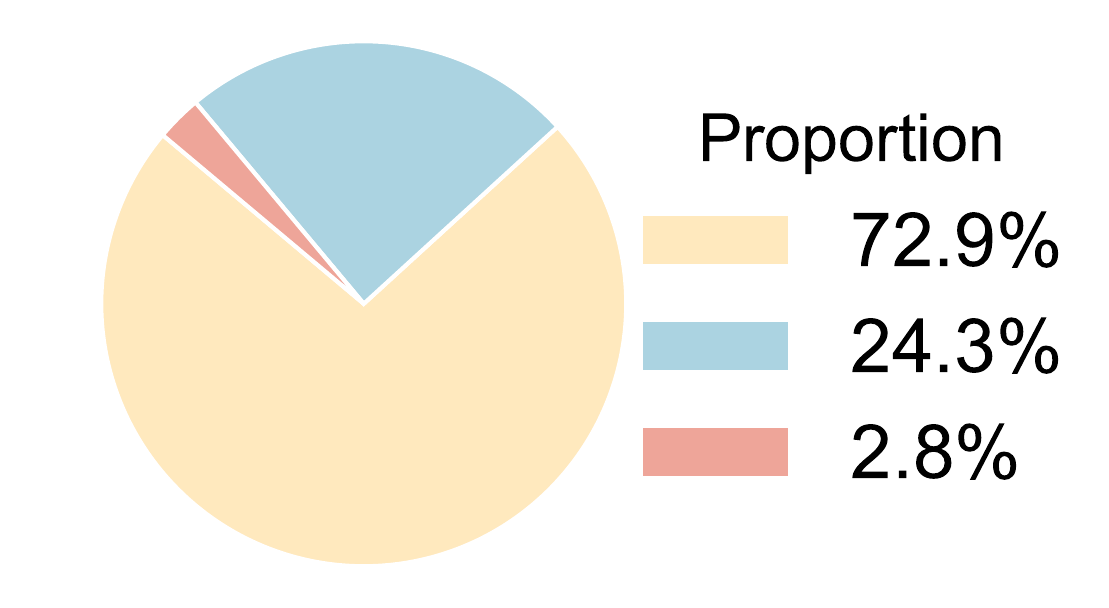}
	}
	
	\caption{Breakdown of Host-to-Device (HtoD) data transfer overhead (Part 1).}
	\label{fig:exp-trans-overhead-part1}
	\vspace{-10pt}
\end{figure}

\subsubsection{Transfer Overhead Analysis}

To further investigate the composition of the data migration bottleneck, we analyze the proportion of transfer time attributed to three main components: Data Vectors (Data), Graph Structure (Graph), and Query Vectors (Query).

As illustrated in Figure~\ref{fig:exp-trans-overhead-part1}, pure GPU-based algorithms, including SONG~\cite{2020SONG}, GGNN~\cite{2022GGNN}, GANNS~\cite{2022GANNS}, and CAGRA~\cite{2024CAGRA}, exhibit nearly identical transfer overhead distributions, as they all require the full migration of both vector data and graph indices. Consequently, we present the SOTA algorithm CAGRA~\cite{2024CAGRA} as the representative for this category, while relegating the results of other pure GPU algorithms to the Figure~\ref{fig:exp-trans-overhead-part2} in Appendix~\ref{ssec:supplementary-of-transfer}. In contrast, hybrid algorithms exhibit distinct profiles, stemming from the divergent hybrid execution strategies employed by BANG~\cite{2024BANG} and PilotANN~\cite{2025PilotANN}. The uniform overhead among pure GPU algorithms indicates that their transfer bottleneck is data-volume dependent. In contrast, the shift in hybrid models like BANG~\cite{2024BANG} and PilotANN~\cite{2025PilotANN} proves that offloading tasks to the CPU can eliminate specific transfer costs, but often relocates the bottleneck or increases synchronization overhead.

\noindent\textit{\textbf{Insight:}} Future systems must optimize the "compute-to-transfer" ratio by compressing the dominant vector data component or aggressively overlapping migration costs with GPU execution.

\subsubsection{Search without Transfer}
To investigate the limits of computational efficiency, we focus on static scenarios where the entire graph structure is permanently resident in GPU memory. This setting isolates the kernel-level search performance, evaluating the raw acceleration capabilities of the search algorithm on the device by excluding data transfer overhead.

As illustrated in Figure~\ref{fig:exp-recall-qps} and Figure~\ref{fig:exp-recall-qps-part2} in Appendix~\ref{ssec:supplementary-of-search-without-trans}, pure GPU-based algorithms achieve over a $10\times$ acceleration compared to traditional CPU baselines. Among them, CAGRA~\cite{2024CAGRA} consistently delivers the highest QPS at 90\% recall on small-scale datasets and maintains relatively stable throughput even as recall targets rise. In contrast, hybrid methods like BANG~\cite{2024BANG} and PilotANN~\cite{2025PilotANN}, despite their scalability, sometimes trail behind even CPU baselines in pure QPS due to the latency incurred by mandatory CPU-side operations. The results highlight that maximizing on-device parallelism is the decisive factor for resident graph search. 

\noindent\textit{\textbf{Insight:}} Future high-performance kernels must prioritize minimizing divergent branching and host synchronization to fully exploit the GPU's massive compute throughput.

\begin{figure}[t]
	\vspace{-15pt}
	\includegraphics[width=0.98\linewidth]{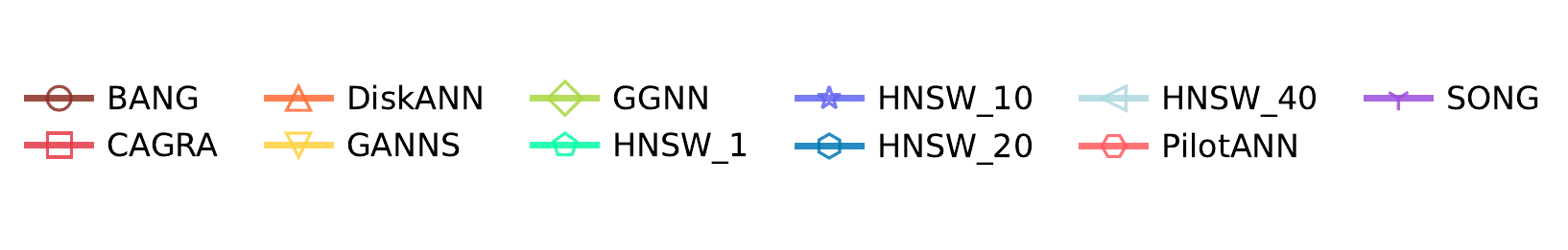}
	\vspace{-10pt}
\end{figure}

\begin{figure}[t]
	\vspace{-10pt}
	\centering
	\subfloat[SIFT1M, batch=10000.]{
		\includegraphics[width=0.47\linewidth]{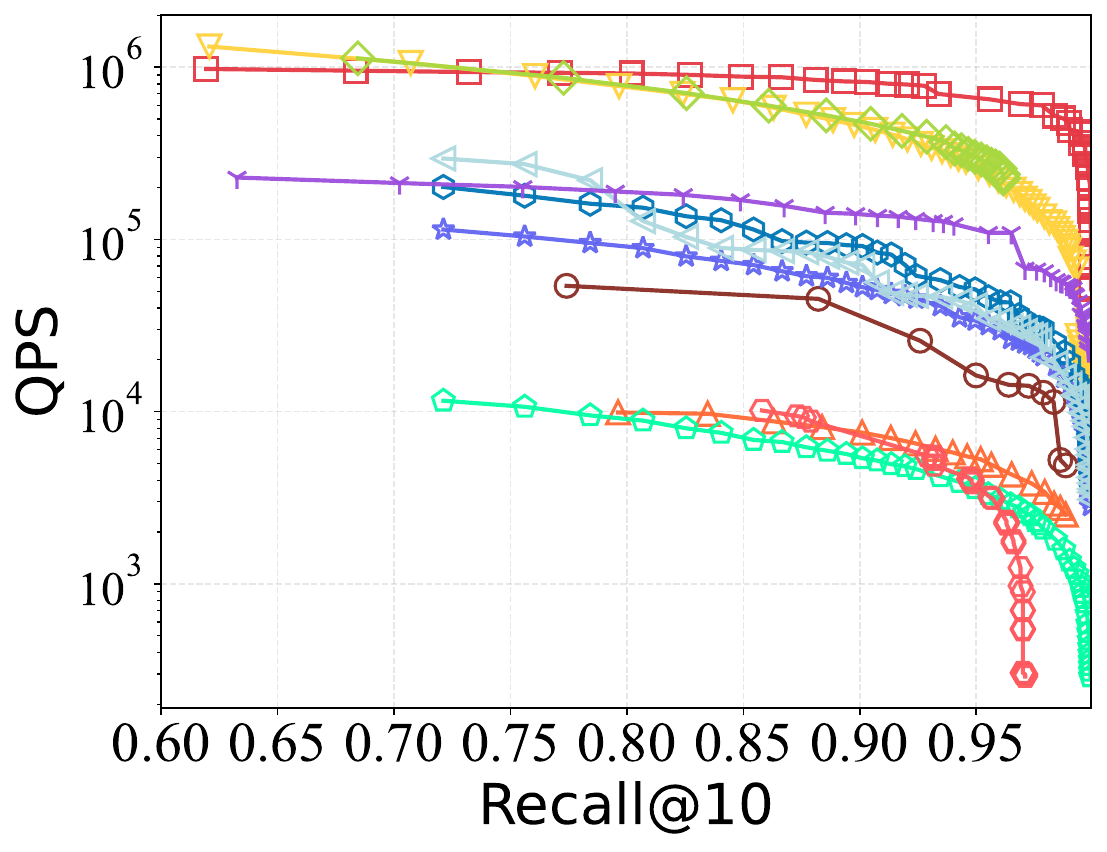}
	}
	\hfill
	\subfloat[DEEP1M, batch=10000.]{
		\includegraphics[width=0.47\linewidth]{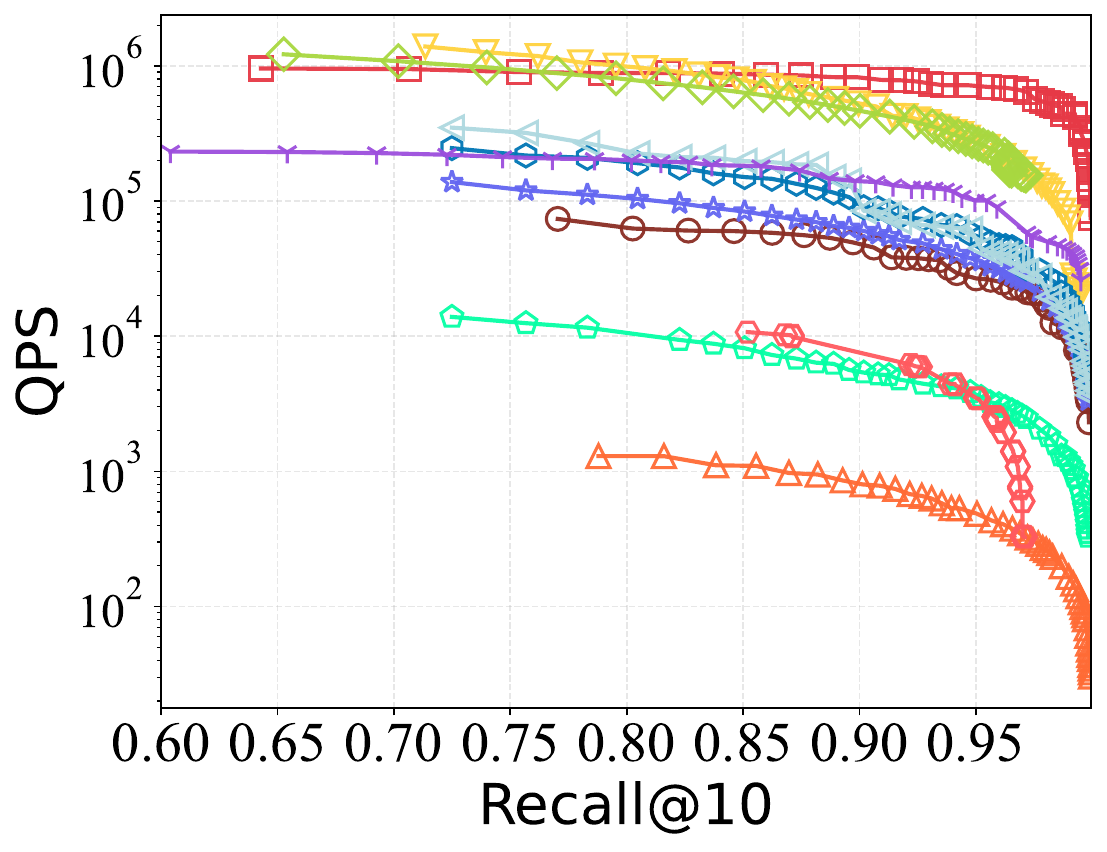}
	}
	\caption{The QPS-Recall@10 of GPU-accelerated graph-based ANNS in high-precision regin (top right is better).}
	\label{fig:exp-recall-qps}
	\vspace{-10pt}
\end{figure}

\begin{figure*}
	\vspace{-10pt}
	\includegraphics[width=0.7\textwidth]{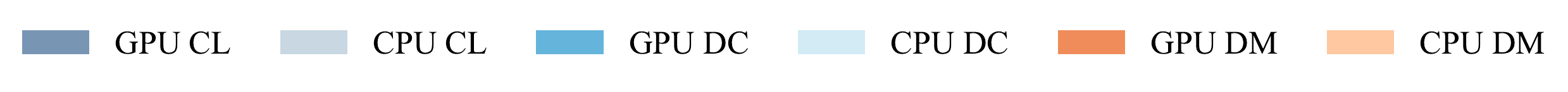}
	\vspace{-10pt}
\end{figure*}

\begin{figure*}[t]
	\vspace{-10pt}
	\centering
	\subfloat[GIST]{
		\includegraphics[width=0.48\linewidth]{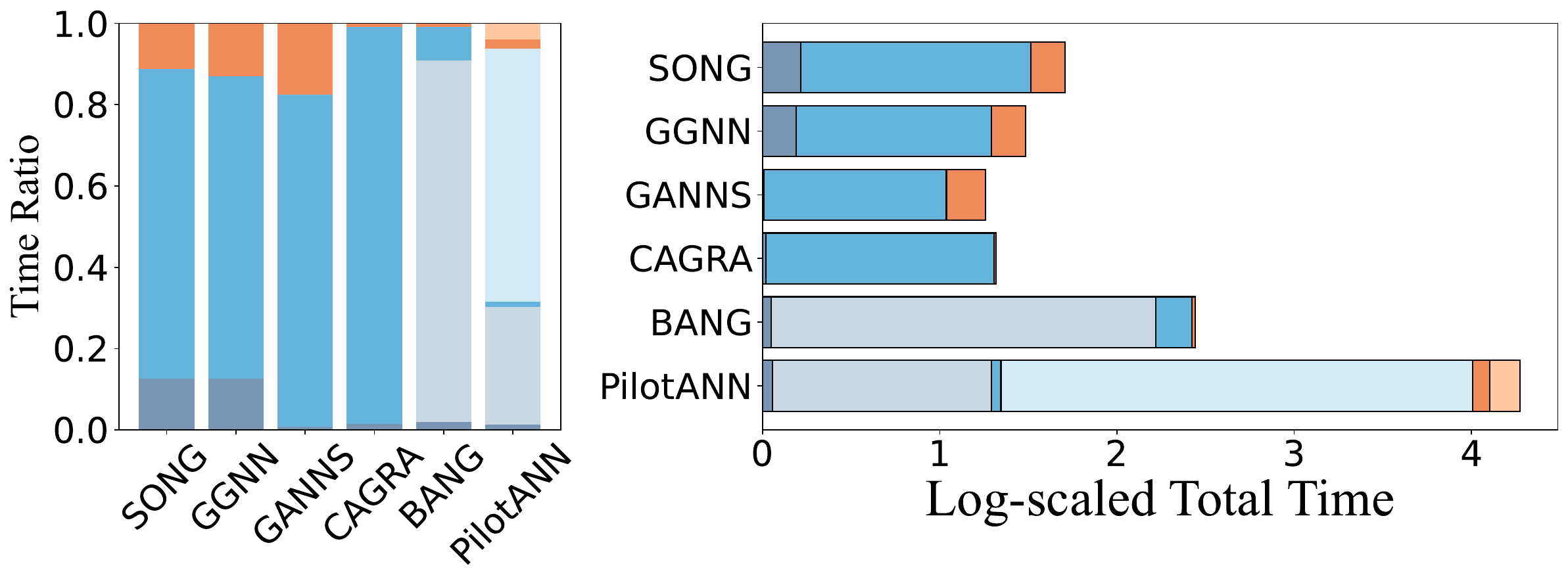}
	}
	\hfill
	\subfloat[DEEP1M]{
		\includegraphics[width=0.48\linewidth]{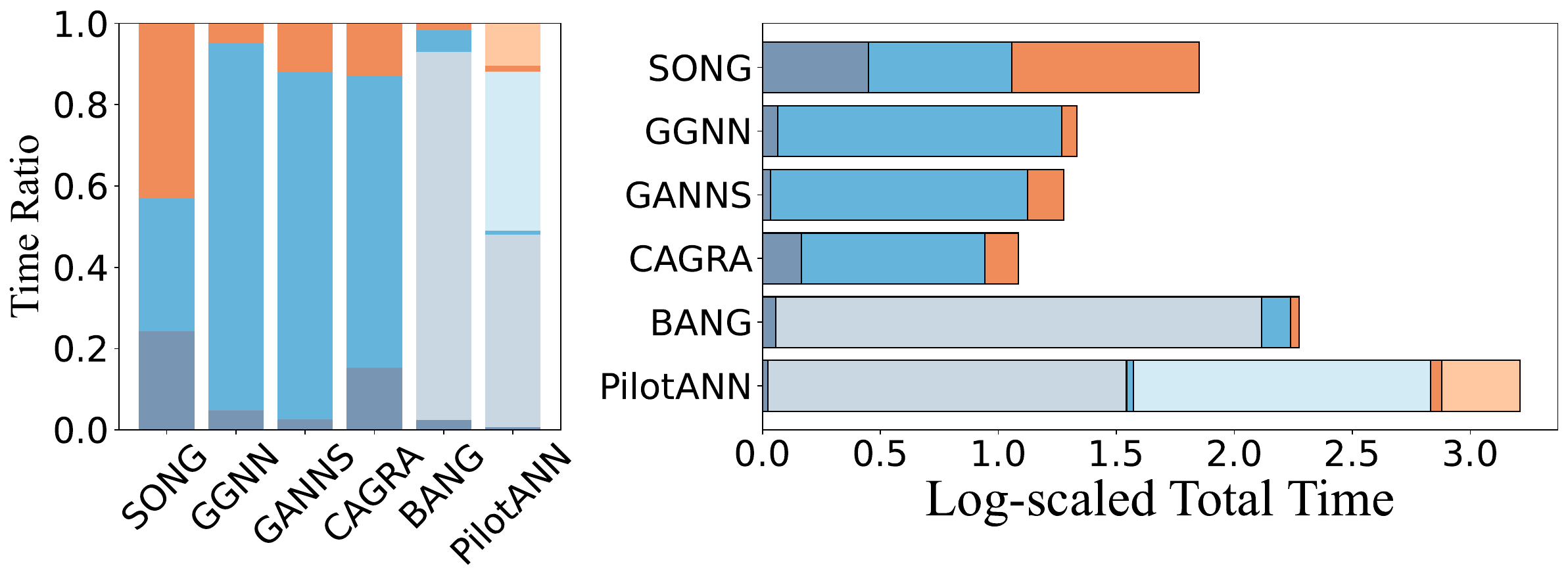}
	}
	\caption{Search component time analysis across different datasets. The charts show the proportion of total kernel time spent on Candidates Locating in Parallel (CL), Bulk Distance Calculation (DC), and Data Structures Maintenance (DM) for various graph-based ANNS algorithms (Part 1).}
	\label{fig:exp-search-components}
\end{figure*}

\subsubsection{Sub-component in Search}
To identify the specific operations that govern search latency, we conduct a fine-grained decomposition of the kernel execution time into three distinct phases: Candidate Locating in Parallel (CL), Bulk Distance Calculation (DC), and Data Structures Maintenance (DM). We analyze the temporal distribution of these components across four datasets with varying characteristics, with all algorithms calibrated to a fixed target of 90\% recall@10.

As illustrated in Figure~\ref{fig:exp-search-components} and Figure~\ref{fig:exp-search-components-part2} in Appendix~\ref{ssec:supplementary-of-subcomp-in-search}, Distance Calculation (DC) consistently dominates the execution profile, accounting for 60-80\% of the total runtime for optimized GPU-native algorithms like GGNN~\cite{2022GGNN}, GANNS~\cite{2022GANNS}, and CAGRA~\cite{2024CAGRA}. While early implementations like SONG~\cite{2020SONG} exhibit a lower relative DC share due to unoptimized overhead in other phases, modern hybrid approaches show distinct patterns: BANG~\cite{2024BANG} reduces the DC proportion to 30-40\% by leveraging Product Quantization (PQ), whereas PilotANN~\cite{2025PilotANN} incurs significant overhead in the Candidate Locating phase due to the latency inherent in CPU-GPU collaboration. While high-dimensional distance calculation is the undisputed bottleneck for algorithms, the results from BANG~\cite{2024BANG} demonstrate that quantization techniques effectively mitigate this computational pressure. More critically, however, in CPU-GPU hybrid architectures, this optimization shifts the bottleneck to the excessive task load on the CPU, where the host becomes overwhelmed by complex search logic and data management.

\noindent\textit{\textbf{Insight:}} Future ultra-fast search kernels must balance arithmetic reduction (via quantization) with optimized host-device workload distribution to prevent CPU saturation from becoming the primary cause of performance degradation.

\begin{figure}[t]
	\centering
	\subfloat[Performance scalability]{
		\includegraphics[width=0.47\linewidth]{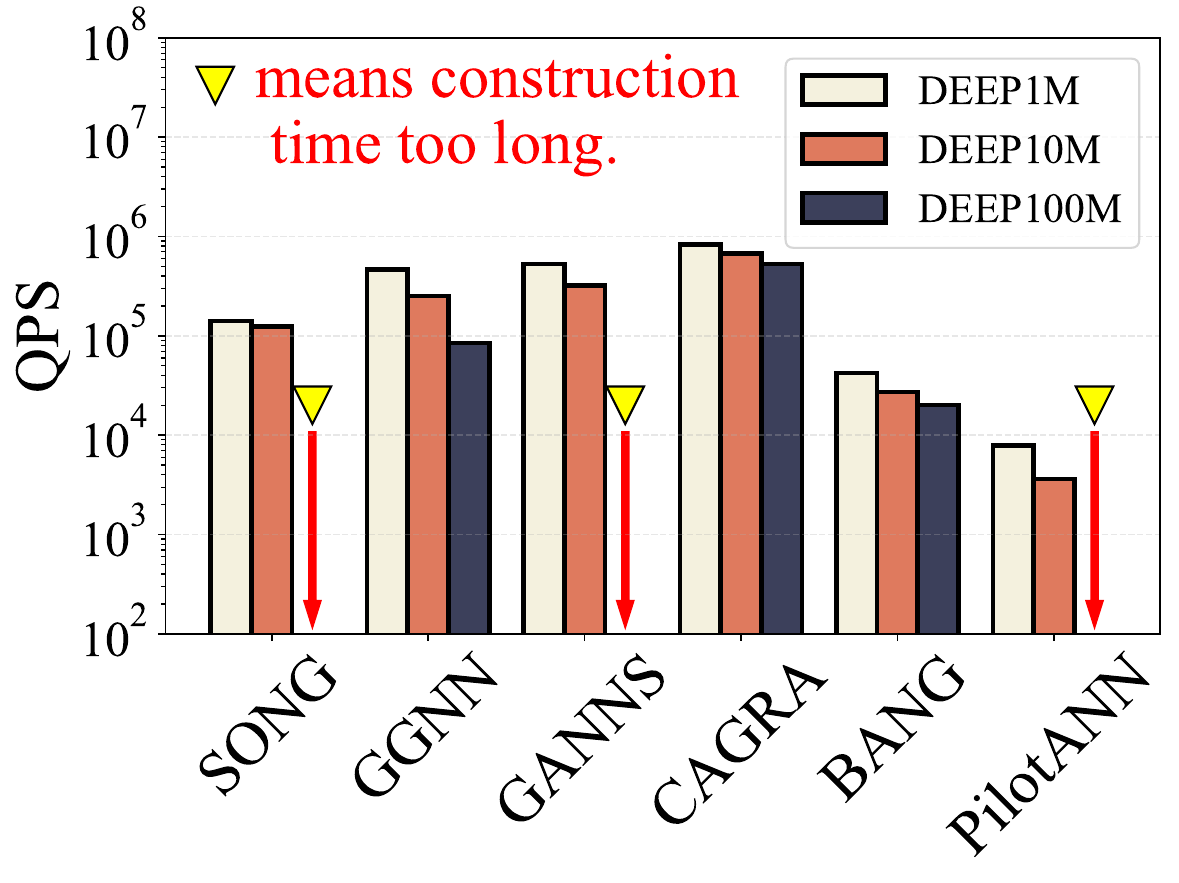}
	}
	\hfill
	\subfloat[Throughput analysis]{
		\includegraphics[width=0.47\linewidth]{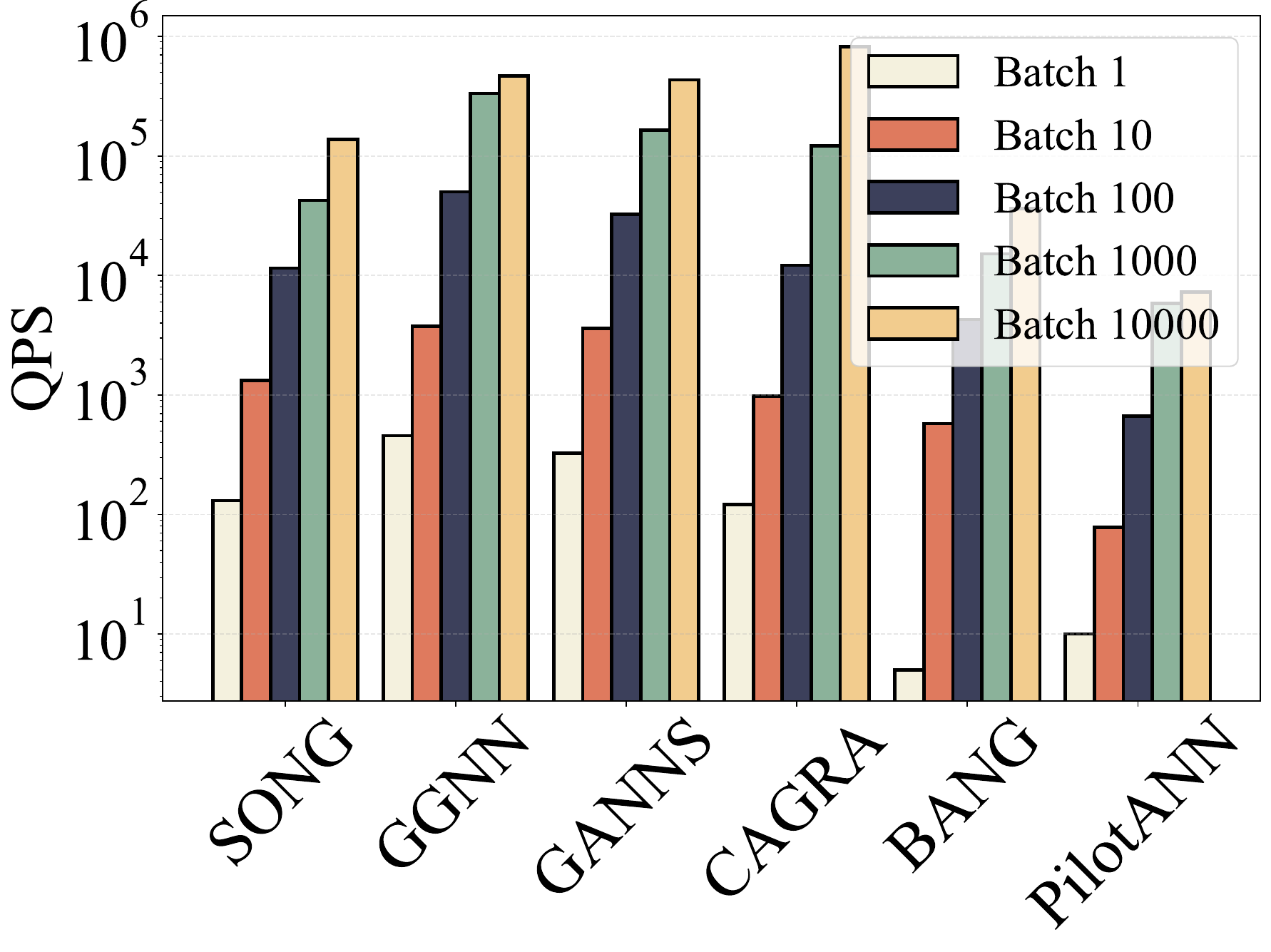}
	}
	\caption{Performance characteristics of GPU-accelerated graph-based ANNS algorithms. (a) Scalability evaluation across DEEP-1M, DEEP-10M, and DEEP-100M datasets at 90\% recall@10. (b) Throughput performance with varying batch sizes on SIFT1M at 90\% recall@10.}
	\label{fig:exp-scalability-and-throught}
	\vspace{-15pt}
\end{figure}

\subsection{Scalability}\label{ssec:scalability}
Specifically, the QPS reported in this section is calculated based on kernel execution time to strictly isolate the algorithmic scalability from data transfer overhead.

\subsubsection{Data Size} We evaluate the scalability of GPU-accelerated algorithms using the DEEP~\cite{DEEP} dataset series, ranging from 1M to 100M vectors. 

As depicted in Figure~\ref{fig:exp-scalability-and-throught}~(a), while performance naturally declines with dataset growth, CAGRA~\cite{2024CAGRA} exhibits the most modest degradation. It maintains high processing speeds even at the 100M scale, demonstrating superior stability compared to competitors that suffer sharp throughput drops due to increased traversal depth and memory pressure. The robust performance of CAGRA~\cite{2024CAGRA} confirms that optimized warp-level parallelism effectively hides the latency spikes associated with larger graph structures.

\noindent\textit{\textbf{Insight:}} Future scalability hinges on the capacity to mask the increasing memory latency inherent in larger graph structures. Research must pivot toward maximizing warp occupancy, ensuring that traversal complexity does not throttle throughput as data volume expands to the billion-scale frontier.

\subsubsection{Throughput Testing}
We evaluate the throughput characteristics of graph-based ANNS by varying the query batch size on the SIFT1M dataset. The experiment measures the resulting QPS while maintaining a fixed Recall@10 of 90\%.

As illustrated in Figure~\ref{fig:exp-scalability-and-throught}~(b), increasing the batch size yields a significant boost in QPS across all algorithms. Since each query is mapped to independent GPU threads, larger batches allow the hardware to hide memory latency more effectively, scaling throughput proportionally with load until hardware resources reach saturation.

\noindent\textit{\textbf{Insight:}} 
Future architectures should prioritize dynamic batching and concurrent execution streams to maximize hardware occupancy. By decoupling efficiency from user-defined batch sizes, systems can consistently saturate thread parallelism and amortize fixed launch overheads regardless of input load.

\section{Guidelines}
Existing GPU-accelerated graph-based ANNS algorithms have clear limitations. SONG~\cite{2020SONG}, as an early method, has limited search optimization and mediocre performance. GGNN~\cite{2022GGNN}, GANNS~\cite{2022GANNS}, and CAGRA~\cite{2024CAGRA} show significant improvements over traditional CPU graph indexing algorithms, but as dataset size grows, these pure GPU algorithms face two critical challenges:

\begin{itemize}[leftmargin=15pt]
 
	\item HtoD transfer overhead surges from 40-80\% on DEEP1M to over 97\% on DEEP100M.
	
	\item GPU memory requirements increase exponentially, severely limiting scalability on large-scale datasets.
\end{itemize}

CPU-GPU hybrid algorithms (BANG~\cite{2024BANG} and PilotANN~\cite{2025PilotANN}) attempt to alleviate memory pressure by performing most search operations on the CPU, thereby reducing the demand on GPU memory. However, this approach merely shifts the bottleneck to the host memory side, as the CPU must handle a large number of distance computations. Consequently, supporting billion-scale datasets still requires massive memory capacity on the host, and the overall performance can be limited by CPU memory bandwidth and latency, rather than GPU computational power.

Therefore, future GPU-accelerated graph-based ANNS algorithms need breakthroughs in two directions:

\begin{itemize}[leftmargin=15pt]
 
	\item For million to ten-million scale datasets, reduce the amount of data transferred to the GPU and lower memory dependency.
	
	\item For billion-scale datasets, reduce host memory consumption while fully exploiting GPU parallelism to achieve higher QPS.
\end{itemize}
}

\section{Conclusion}\label{sec:conclusion}

 
 
 This survey comprehensively analyzes graph-based GPU approximate nearest neighbor search algorithms by evaluating 6 GPU algorithms and 2 CPU algorithms across 8 datasets ranging from 0.29M to 100M in scale, examining search performance (QPS, query path length, distance computations, batch throughput, and parameter sensitivity) and construction efficiency (build time, memory, and graph connectivity). The study finds that distance calculations dominate search overhead and, for the first time, quantifies CPU-GPU transfer costs, showing data transfer time far exceeds kernel execution time. We propose using average query path length as a graph quality metric that better reflects navigation efficiency, providing clear directions for algorithm optimization.
 
 \clearpage
 
\balance
\bibliographystyle{ACM-Reference-Format}
\bibliography{vector_search}

\clearpage
\appendix
\section{Appendix}

\subsection{GPU Architectures}\label{ssec:gpu-arch}

\noindent{\textbf{Architecture Overview.}} NVIDIA's recent architectures have evolved from Ampere and Hopper to Blackwell, with progressive improvements in memory and communication bandwidth. Taking Ampere as an example, as shown in Nvidia report~\cite{nvidia2020ampere}, the architecture features a hierarchical design with multiple Graphics Processing Clusters (GPCs), each containing Texture Processing Clusters (TPCs) and Streaming Multiprocessors (SMs). A typical Ampere GPU has dozens of SMs, each housing multiple CUDA cores, tensor cores, and shared memory units. This design enables thousands of concurrent threads, making it well-suited for the parallelism in ANNS.

\begin{figure}[h]
	\centering
	\includegraphics[width=0.9\linewidth]{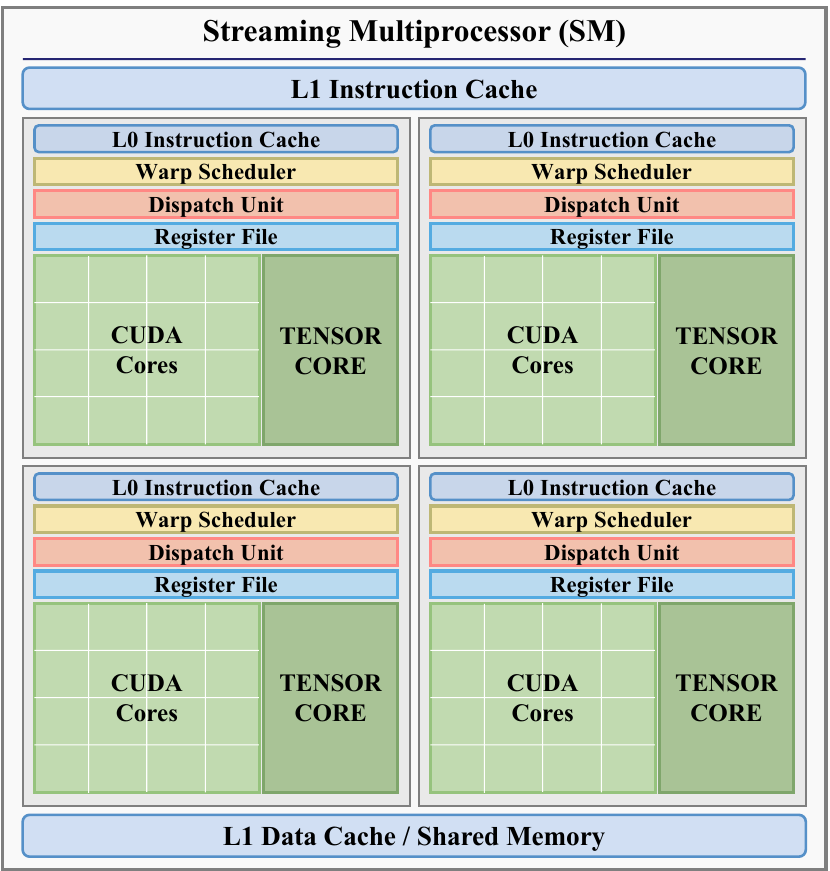}
	\caption{Internal architecture of a Streaming Multiprocessor (SM), based on the Ampere architecture~\cite{nvidia2020ampere}.}
	\label{fig:sm-arch}
\end{figure}

\noindent{\textbf{Streaming Multiprocessor (SM).}} Each SM serves as the fundamental execution unit, containing 128 CUDA cores, 4 tensor cores, and 128KB of configurable shared memory. The SM employs a SIMT~\cite{nickolls2008scalable} execution model, where threads are organized into warps of 32 threads each. Within an SM, multiple warps can be active simultaneously, with the warp scheduler dynamically selecting ready warps for execution. The internal architecture of an SM is illustrated in Figure~\ref{fig:sm-arch}.

\subsection{Graph Types}\label{ssec:graph-type}
According to a previous survey~\cite{wang2021comprehensive}, existing graph-based ANNS algorithms are fundamentally derivatives of these four base graphs. \textcolor{black}{In particular, due to the poor connectivity of MSTs and the resulting long average search paths, they are used solely to ensure basic graph connectivity and do not serve as the final graph.}

\textbf{Delaunay Graph (DG).} Within Euclidean space $\mathbb{R}^d$, a Delaunay Graph $G(V, E)$ constructed on $\mathcal{S}$ must satisfy the following condition: for any edge $e \in E$ connecting vertices $u, v \in V$, there exists a circle passing through both $u$ and $v$ such that no other vertices lie strictly inside this circle, with at most three vertices permitted on the circle's boundary.

\textbf{Relative Neighborhood Graph (RNG).} A Relative Neighborhood Graph $G(V, E)$ constructed on a point set $\mathcal{S}$ exhibits the following property: for any edge $e \in E$ connecting vertices $u, v \in V$, it must hold that for every other vertex $w \in V \setminus \{u, v\}$, either $\text{dist}(u,v) < \text{dist}(u,w)$ or $\text{dist}(u,v) < \text{dist}(v,w)$.

\textbf{K-Nearest Neighbor Graph (KNNG).} The construction of KNNG $G(V, E)$ involves connecting every point in $\mathcal{S}$ to its $k$ closest neighbors. Within this structure, for any vertices $u, v \in V$, the relationship $u \in \mathcal{N}(v)$ does not guarantee that $v \in \mathcal{N}(u)$.

\textbf{Minimum Spanning Tree (MST).} An MST represents the graph $G(V, E)$ that achieves the minimal value of $\sum_{i=1}^{|E|} w(e_i)$ across dataset $\mathcal{S}$, where each edge $e_i$ establishes a connection between vertices $u$ and $v$, with $w(e_i) = \text{dist}(u,v)$.

\subsection{Algorithm of Graph Search on GPU}\label{ssec:alg-of-graph-search}

\begin{algorithm}[h]
	\small
	\caption{Standardized Greedy Search for Graph-Based ANNS on GPU Architecture}
	\label{alg:greedy-search}
	
	\KwIn{Graph $G=(V,E)$, Query Batch $\mathcal{Q} = \{q_1, q_2, \ldots, q_m\}$}
	\KwOut{Top-$k$ nearest neighbors $\mathcal{R} = \{r_1, r_2, \ldots, r_m\}$}
	
	\tcp{Randomly dispatch $\mathcal{Q}$ into different thread blocks}
	
	\For{\textbf{each} $q_i \in \mathcal{Q}$ \textbf{parallel across blocks}}{
		$C \gets \emptyset, V_{visited} \gets \emptyset$ \tcp*[r]{Candidates, Visited Table}
		$r_i \gets \emptyset$ \tcp*[r]{Result Queue on $Block_i$}
		$C.push(v_{entry}, \text{dist}(v_{entry}, q_i))$\;
		
		\While{$C \neq \emptyset$}{
			$v_{current} \gets C.popMin()$\;
			\If{termination condition met}{
				\textbf{break} \tcp*[r]{e.g., Iteration Limit}
			}
			$r_i.insert(v_{current}), V_{visited}.update(v_{current})$\;
			
			\tcp{Parallel neighbor filtering within block}
			\For{\textbf{each} $u \in \mathcal{N}(v_{current})$ \textbf{parallel}}{
				\If{$u \notin V_{visited}$}{
					$n_{unvisited}.push(u)$\;
				}
			}
			
			\tcp{Parallel distance calculation}
			\For{\textbf{each} $n \in n_{unvisited}$ \textbf{parallel}}{
				$d_{unvisited}.push(n, \text{dist}(n, q_i))$\;
			}
			
			$C \gets \text{Sort}(\text{Merge}(C, d_{unvisited}))$\;
		}
		$r_i \gets r_i.top(k)$\;
	}
	
	\tcp{Merge results from all blocks}
	\Return $\mathcal{R} \gets \bigcup_{i=1}^{m} r_i$\;
\end{algorithm}

\begin{table*}[t]
	\centering
	\small
	\caption{Index construction metrics on medium to large-scale datasets (Part 2). For the MNIST8M and DEEP100M datasets,  "OOM$^\dagger$  " and "OOM$^{*}$  " denote experiments omitted due to exceeding the system memory and GPU memory limits, respectively.}
	\setlength{\tabcolsep}{2pt}
	\renewcommand{\arraystretch}{1.2}
	\begin{tabular}{c|cccc|cccc|cccc}
		\hline
		\multirow{2}{*}{Alg} 
		& \multicolumn{4}{c|}{DEEP10M~\cite{DEEP}}   
		& \multicolumn{4}{c|}{MNIST8M~\cite{lecun1998mnist}}   
		& \multicolumn{4}{c}{DEEP100M~\cite{DEEP}} \\
		\cline{2-13}
		& CT & AD & CC & PMF 
		& CT & AD & CC & PMF 
		& CT & AD & CC & PMF  \\
		\hline
		SONG~\cite{2020SONG}     &33413s   & 47  & 1199  &13557/0   &  123031s & 104  & 15  & 40126/0  & OOM$^\dagger$    & OOM$^\dagger$    &OOM$^\dagger$    & OOM$^\dagger$    \\
		
		GGNN~\cite{2022GGNN}    & \underline{293s}  & 64  &  24 & 6272/9104  & 3125s  & 128  & 5  & 25837/46867  & \textbf{2065s$^{*}$}  & 32  & 1  & 49100/76237   \\
		
		GANNS~\cite{2022GANNS}      & 1488s  & 48  & 1074 & 6272/9104 & 12448s  & 101  & 7  & 32225/45527  & OOM$^{*}$     & OOM$^{*}$      & OOM$^{*}$    & OOM$^{*}$    \\
		
		CAGRA~\cite{2024CAGRA}  & \textbf{102s$^{*}$}  & 64  & 1  & 37144/12441  & \textbf{1238s$^{*}$}  & 128  & 1  & 28447/31679  & \underline{9031s}  & 32  & 1  & 87124/43237  \\
		
		Pilot-ANN~\cite{2025PilotANN} & 11529s  & 45  & 1  & 15563/0  & 25340s  & 56  & 1  & 75432/0   &OOM$^\dagger$     &OOM$^\dagger$     & OOM$^\dagger$   &OOM$^\dagger$     \\
		
		HNSW~\cite{2018HNSW}      & 536s  & 46  & 1  & 12580/0  & \underline{2205s} & 86  & 1  & 56601/0 &OOM$^\dagger$    &OOM$^\dagger$    &OOM$^\dagger$    &OOM$^\dagger$    \\
		
		VAMANA~\cite{2019diskann}   &  4903s &  64 & 1  & 12781/0  &14991s   & 128  & 1  &59820/0   & 24086s  &  32 & 1  & 60650/0  \\
		\hline
	\end{tabular}
	\label{tab:construction-eff-part2}
\end{table*}

\subsection{Impact of Hyper Parameters}\label{parameters}

\begin{figure}[h]
	\vspace{-15pt}
	\centering
	\subfloat[Impact of search width.]{
		\includegraphics[width=0.47\linewidth]{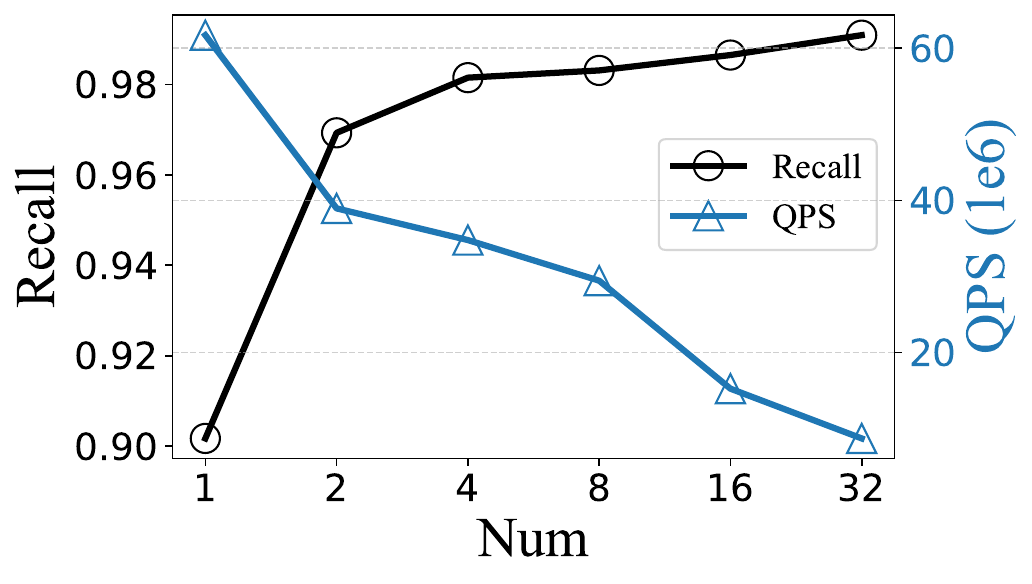}
	}
	\hfill
	\subfloat[Impact of internal topk size.]{
		\includegraphics[width=0.47\linewidth]{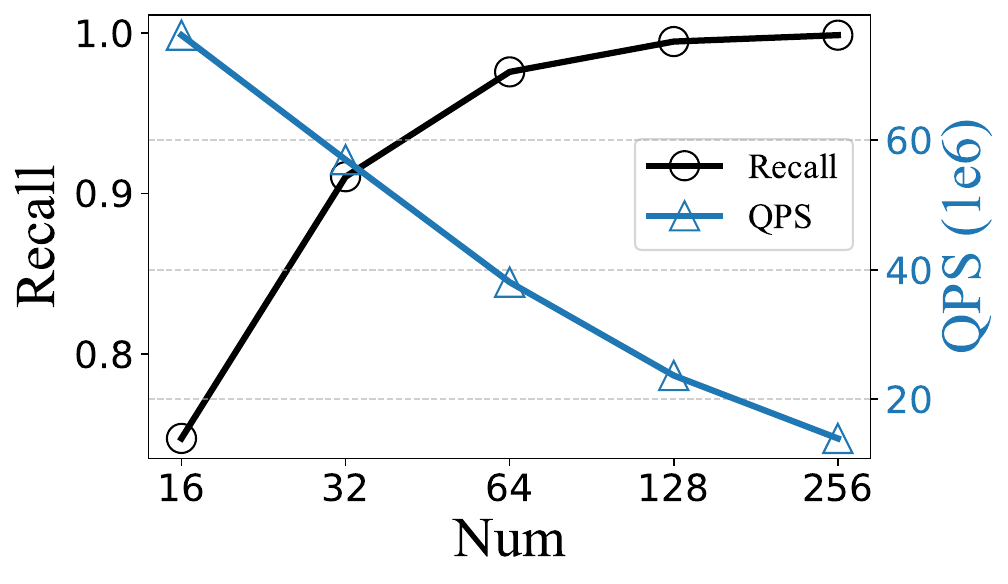}
	}
	\hfill
	\subfloat[Impact of graph degree.]{
		\includegraphics[width=0.47\linewidth]{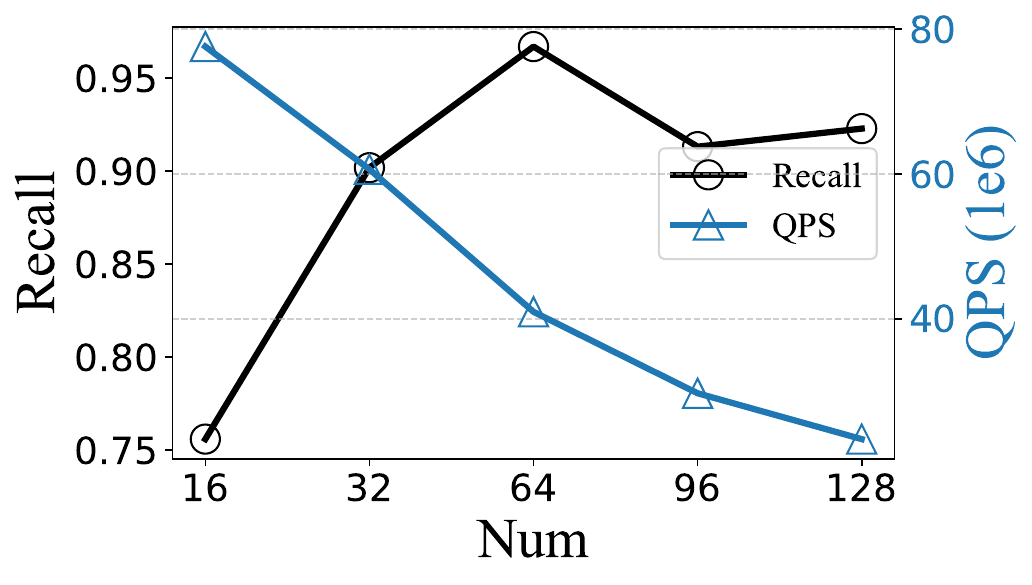}
	}
	\hfill
	\subfloat[Impact of team size.]{
		\includegraphics[width=0.47\linewidth]{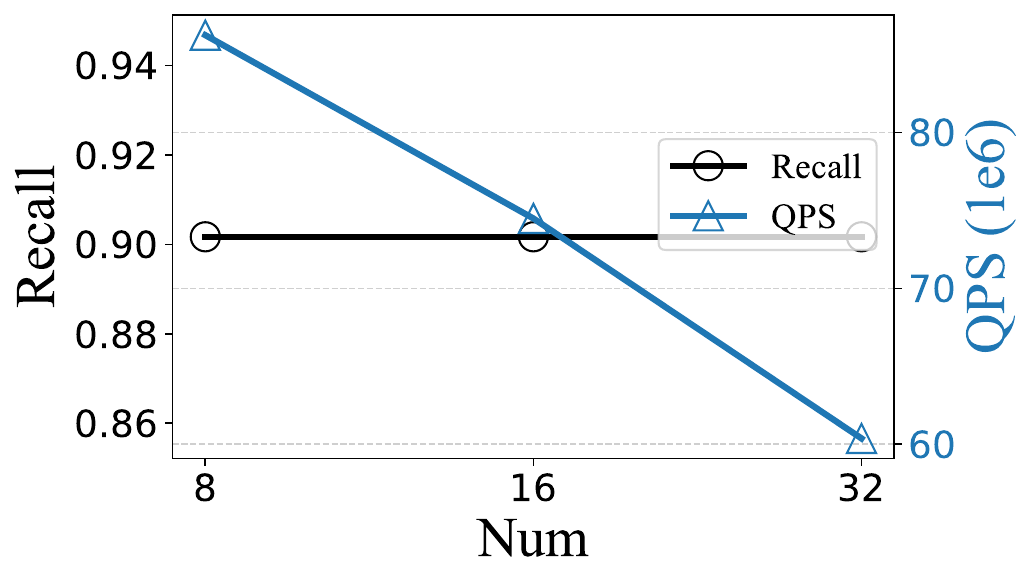}
	}
	\caption{Impact of different hyper parameters in CAGRA.}
	\label{fig:hyper-parameters}
	\vspace{-10pt}
\end{figure}

We select the algorithm with the highest and the stablest QPS, CAGRA~\cite{2024CAGRA}, to analyze the impact of hyper parameters.We conducted a systematic evaluation of the CAGRA algorithm, measuring its QPS and recall under different hyper parameter settings. Specifically, we examined the effects of the degree parameter, the internal top‑k size, search width, and team size on performance. The experiments are launched on the SIFT~\cite{IVF} The result can be seen in Figure~\ref{fig:hyper-parameters}:

\observation{} When tuning search width and the internal top-k, QPS and recall tend to move in opposite directions. A high out-degree doesn't necessarily lead to better search performance. Team size doesn't affect recall, but it can speed up the search.

\analysis In graph-based search, the effects of parameters on the algorithm are complex and interdependent. During the build phase, increasing the degree may reduce the navigability of the graph, as additional edges do not always contribute to efficient traversal. As a result, the search may need to visit more points before approaching the query, potentially increasing the computational cost.

\begin{figure*}[t]
	\vspace{-10pt}
	\includegraphics[width=0.98\linewidth]{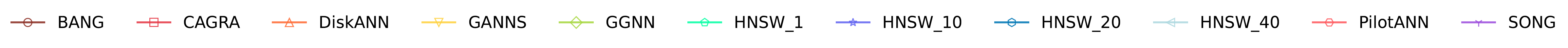}
	\vspace{-10pt}
\end{figure*}

\begin{figure*}[t]
	\vspace{-10pt}
	\centering
	\subfloat[NYTimes, batch=10000.]{
		\includegraphics[width=0.28\linewidth]{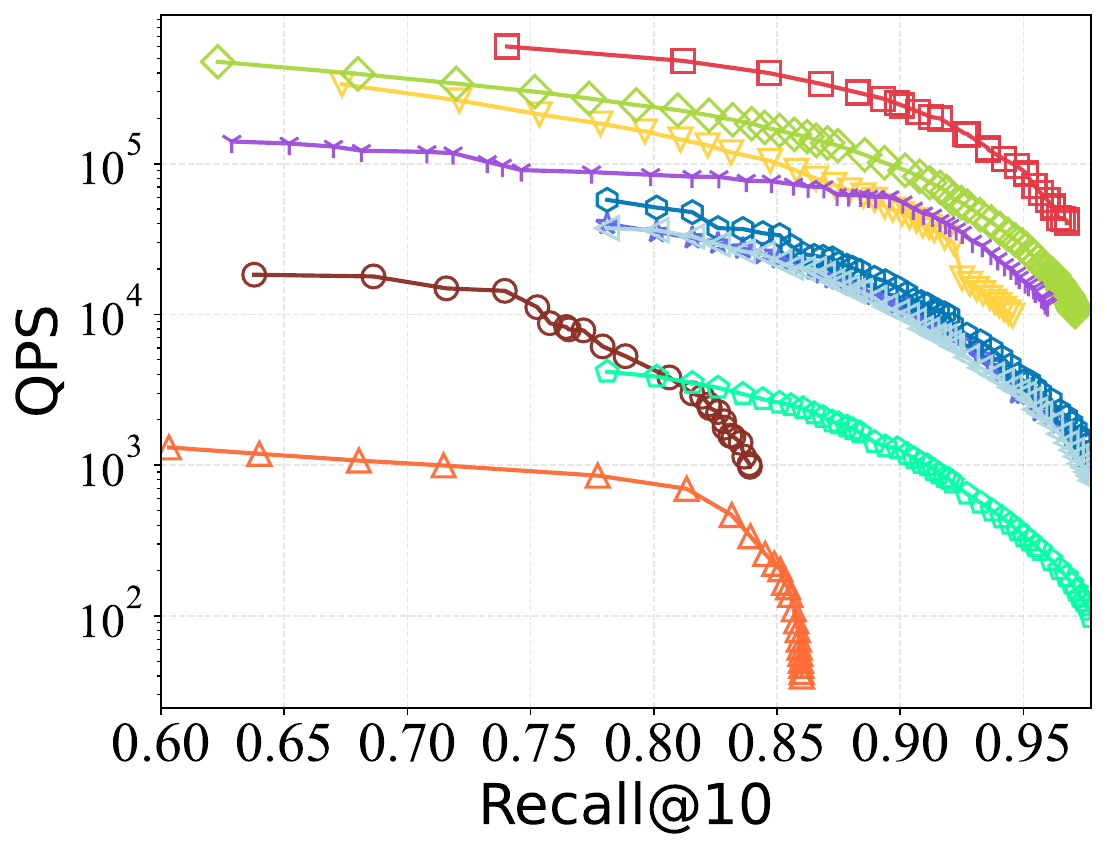}
	}
	\hfill
	\subfloat[GLOVE, batch=10000.]{
		\includegraphics[width=0.28\linewidth]{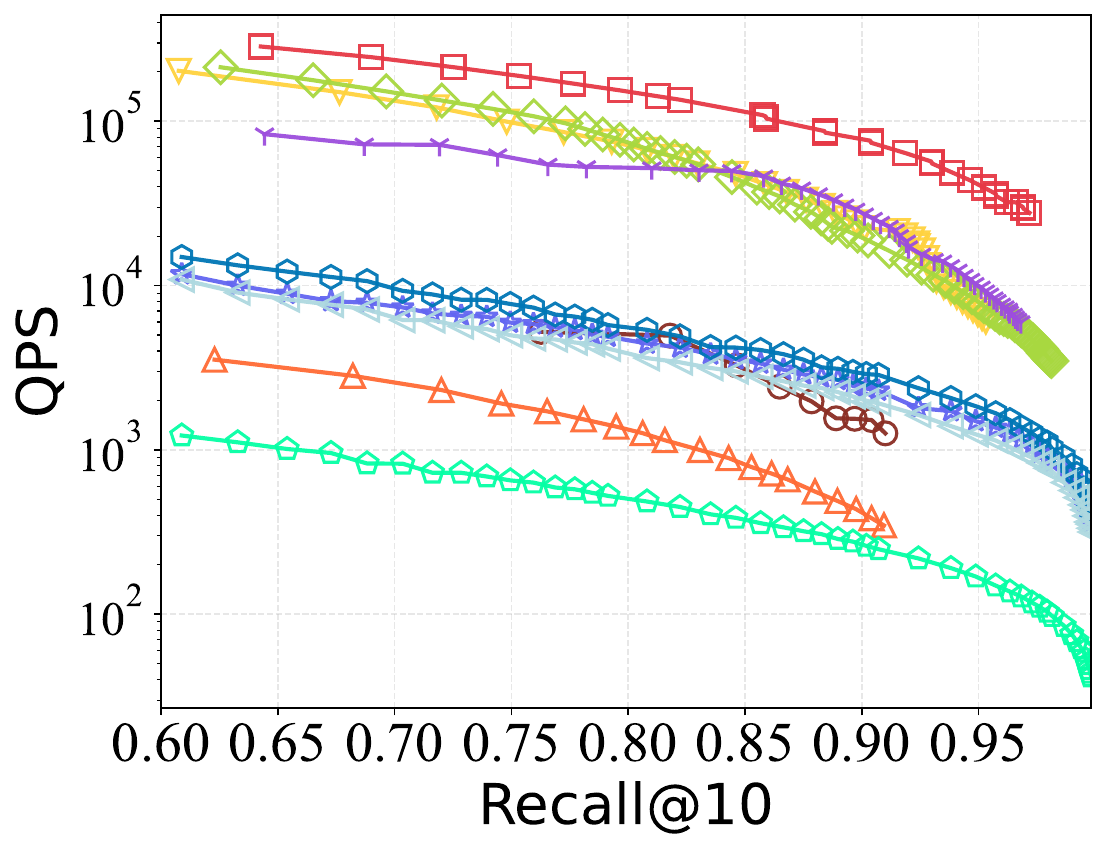}
	}
	\hfill
	\subfloat[GIST, batch=1000.]{
		\includegraphics[width=0.28\linewidth]{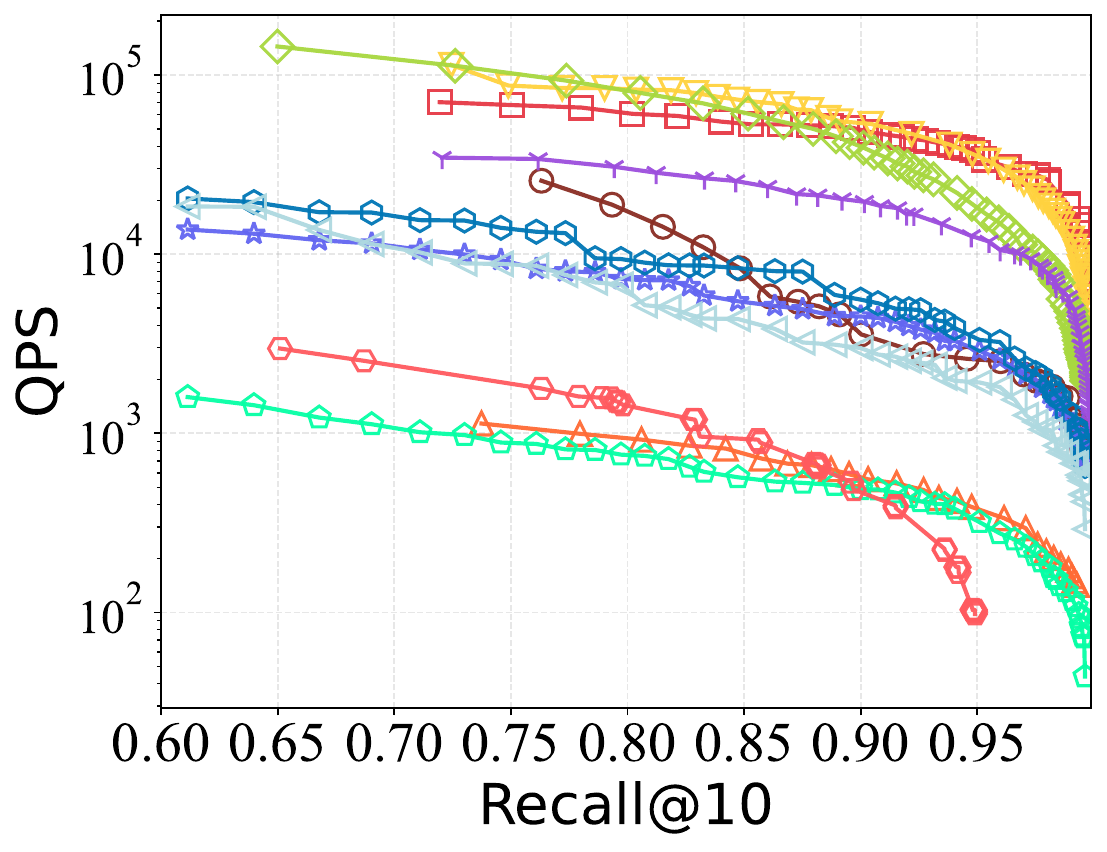}
	}
	\hfill
	\subfloat[DEEP10M, batch=10000.]{
		\includegraphics[width=0.28\linewidth]{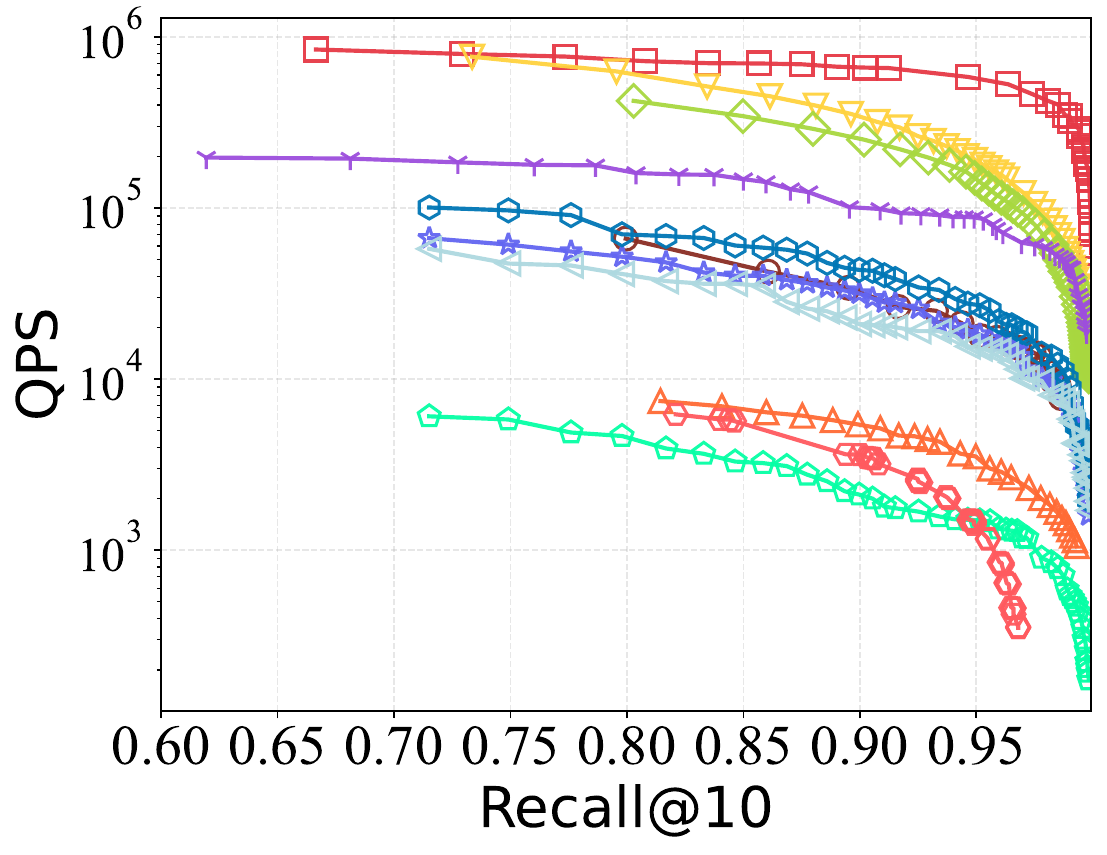}
	}
	\hfill
	\subfloat[MNIST8M, batch=10000.]{
		\includegraphics[width=0.28\linewidth]{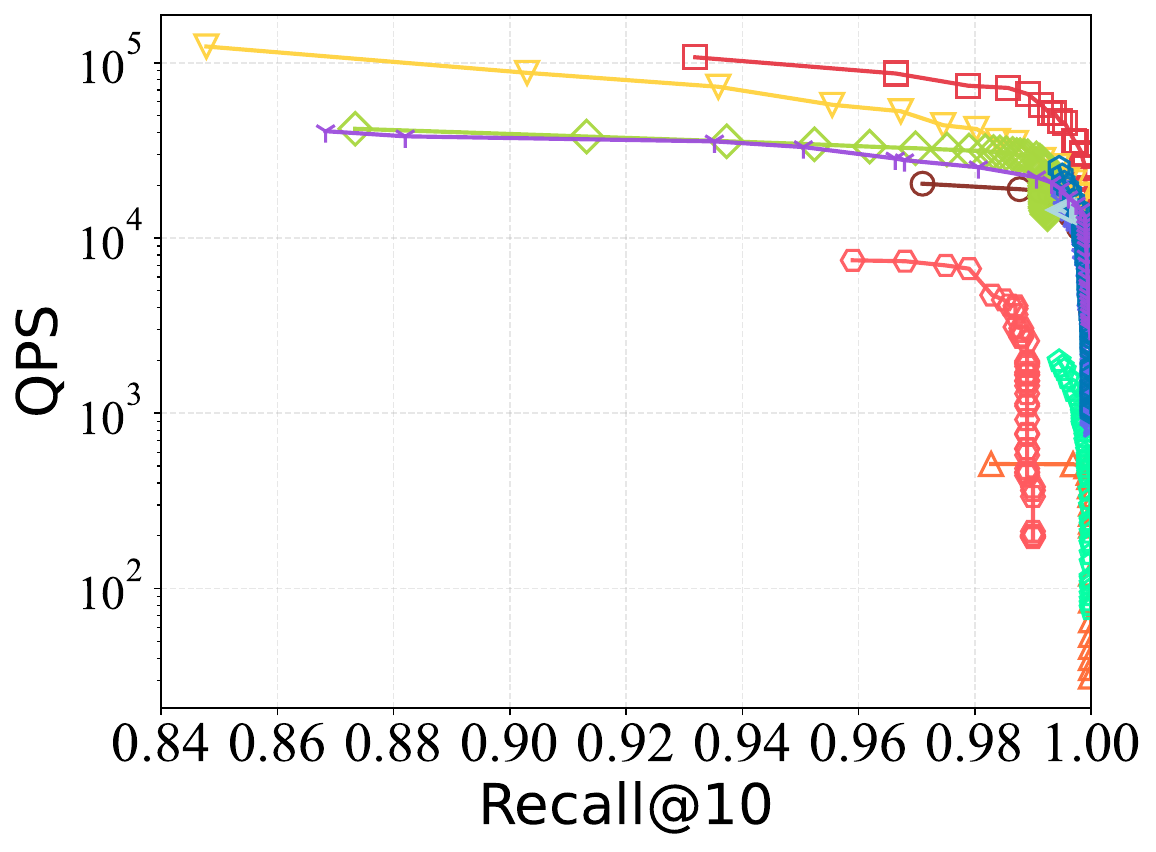}
	}
	\hfill
	\subfloat[DEEP100M, batch=10000.]{
		\includegraphics[width=0.28\linewidth]{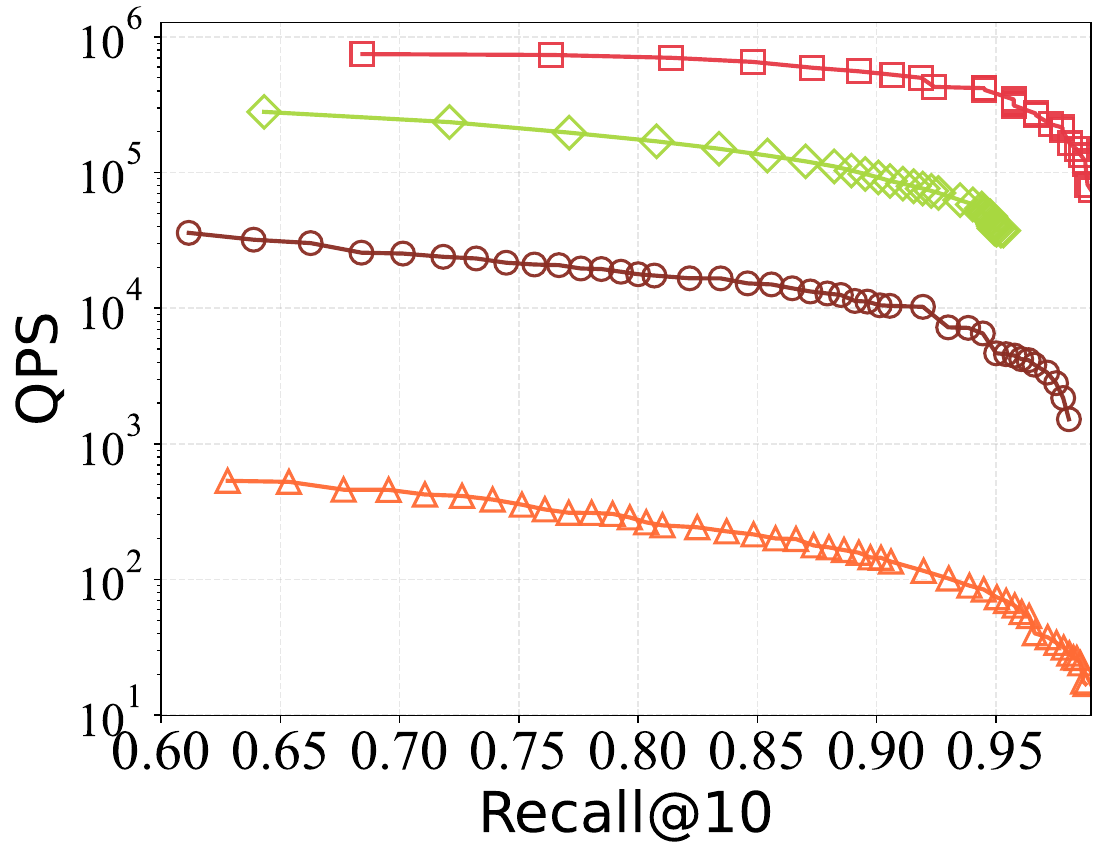}
	}
	\caption{The QPS-Recall@10 of GPU-accelerated graph-based ANNS in high-precision regin (Part 2).}
	\label{fig:exp-recall-qps-part2}
\end{figure*}

\subsection{Graph-based ANNS Algorithms on GPU Architectures}\label{ssec:graph-anns-gpu}
We present an overview of the baseline algorithms as follows:


\noindent{\textbf{A1: SONG~\cite{2020SONG}.}} SONG is the first GPU-specific graph-based ANNS algorithm, achieving efficient search through architectural optimization. It decouples traditional graph search into three independent stages: candidate locating, bulk distance computation, and data structure maintenance, eliminating sequential dependencies. The algorithm employs fixed-degree adjacency lists, open addressing hash tables, and Bloom filters to avoid dynamic memory allocation and optimize access patterns. Through bounded priority queues and selective insertion mechanisms, it achieves fine-grained GPU memory control while maintaining sequential access table updates during parallel distance computation.


\textcolor{black}{\noindent{\textbf{A2: GGNN~\cite{2022GGNN}.}} GGNN employs a bottom-up hierarchical graph construction method to enhance both index building efficiency and query performance. The construction process partitions the dataset into small batches, recursively merges subgraphs layer by layer, and strengthens connectivity through symmetric linking while maintaining a fixed number of outgoing edges per node. During search, GGNN utilizes thread-block-level parallelism with a multi-purpose shared-memory cache and an adaptive stopping criterion to rapidly locate nearest neighbors. For billion-scale datasets, GGNN supports a multi-GPU sharding architecture. While each shard is constructed as an independent search-graph to fit within individual GPU memory, a unified query is executed across all shards, and their respective results are aggregated on the CPU via an efficient n-way merge to provide the final global k-nearest neighbors.}

\noindent{\textbf{A3: GANNS~\cite{2022GANNS}.}} GANNS is an efficient GPU-accelerated algorithm based on proximity graphs, utilizing GPU acceleration in both the graph construction and search processes. During the graph construction phase, GANNS adopts a divide-and-conquer strategy, dividing the point set into multiple subsets to construct local proximity graphs in parallel. These local graphs are then gradually merged to form a global graph, while maintaining forward and backward edges to ensure the graph's high quality. In the search phase, GANNS processes candidate points and distance calculations in batches, combining lazy updates and lazy checks to avoid the overhead of dynamic data structures while enhancing parallelism.

\noindent{\textbf{A4: CAGRA~\cite{2024CAGRA}.}} CAGRA is a nearest neighbor algorithm that uses GPU acceleration for graph construction and search. In terms of graph construction, CAGRA employs the NN-Descent~\cite{nn-descent} algorithm to quickly generate an initial k-nearest neighbor graph, followed by optimization through edge reordering and reverse edge addition to enhance graph connectivity and search performance. The optimized graph features a fixed out-degree, effectively leveraging the parallel computing capabilities of GPUs and improving hardware resource utilization. In terms of graph search, CAGRA initializes the candidate node list through random sampling and achieves efficient iterative search by dynamically updating the internal Top-M list and the candidate list.

\noindent{\textbf{A5: BANG~\cite{2024BANG}.}} BANG is a GPU-accelerated ANNS method designed for billion-scale datasets. It constructs the graph using the classic DiskANN~\cite{2019diskann} algorithm. During the search phase, BANG reduces GPU memory usage by storing the graph structure and raw data in host memory. The CPU fetches neighbor nodes and sequentially transfers their IDs and corresponding data to the GPU. On the GPU side, BANG employs a phased execution strategy that decomposes the workflow into neighbor filtering, waiting for neighbor data arrival, distance computation, prefetching expansion candidates, and re-ranking. By leveraging asynchronous data transfers and data prefetching, BANG maintains parallelism while effectively mitigating the CPU–GPU data transfer bottleneck.


\noindent{\textbf{A6: PilotANN~\cite{2025PilotANN}.}} PilotANN is a hybrid CPU-GPU framework for large-scale ANNS. During indexing, it applies Singular Value Decomposition (SVD)~\cite{klema1980svd, golub2013matrix} to split high-dimensional vectors into primary and residual components, constructs subgraphs through node sampling, and uses Compressed Sparse Row (CSR)~\cite{saad2003iterative} format to fit the subgraph in GPU memory while keeping the full graph in CPU memory. During search, it employs a multi-stage strategy: an initial GPU-based search using the subgraph and primary vectors identifies candidates, CPU refinement incorporates residual vectors and further subgraph traversal, and a final greedy search on the full graph ensures precision. Through Fast Entry Selection (FES) and efficient data management, PilotANN achieves high throughput for ultra-large-scale datasets.

\subsection{Supplementary of Construction Evaluation}\label{ssec:supplementary-of-construction}

Table~\ref{tab:construction-eff-part2} supplements the main evaluation by detailing construction performance on medium-to-large datasets (DEEP10M, MNIST8M, and DEEP100M). It highlights the scalability limits of different algorithms, where CPU-based methods like HNSW~\cite{2018HNSW} and Pilot-ANN~\cite{2025PilotANN} frequently encounter system memory exhaustion (OOM$^\dagger$) on the largest scale, while GPU-native solutions like CAGRA~\cite{2024CAGRA} and GGNN~\cite{2022GGNN} maintain robust throughput.

\subsection{Supplementary of Transfer Overhead Analysis}\label{ssec:supplementary-of-transfer}

\begin{figure}[h]
	\vspace{0pt}
	\includegraphics[width=0.7\linewidth]{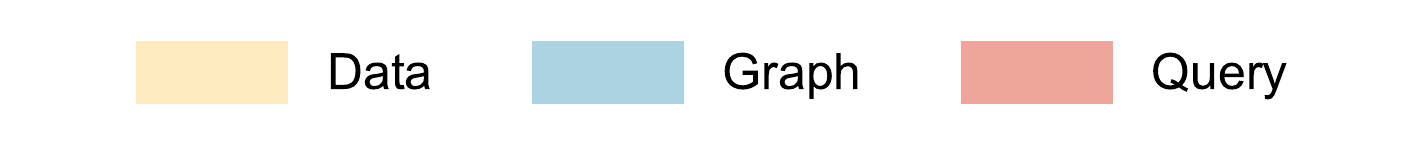}
	\vspace{-10pt}
\end{figure}

\begin{figure}[h]
	\vspace{-10pt}
	\centering
	\subfloat[GGNN.]{
		\includegraphics[width=0.31\linewidth]{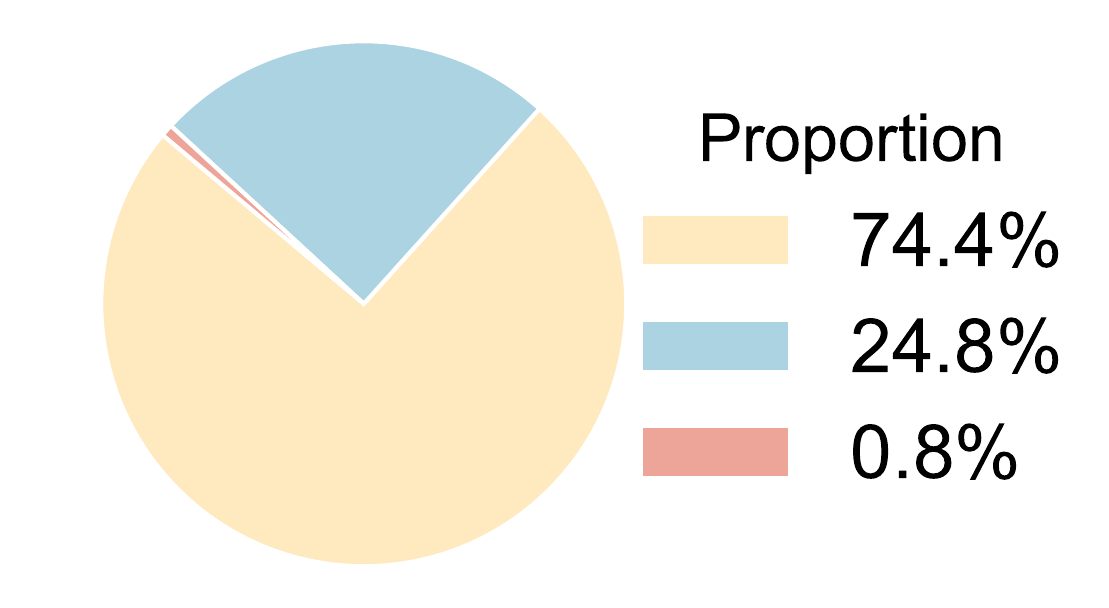}
	}
	\hfill
	\subfloat[GANNS.]{
		\includegraphics[width=0.31\linewidth]{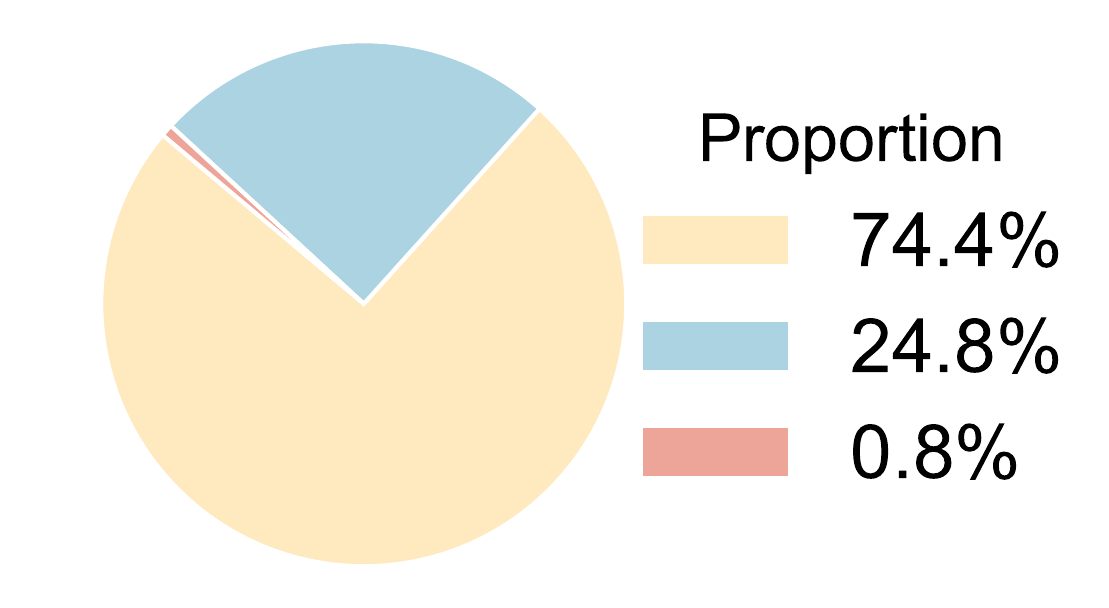}
	}
	\hfill
	\subfloat[SONG.]{
		\includegraphics[width=0.31\linewidth]{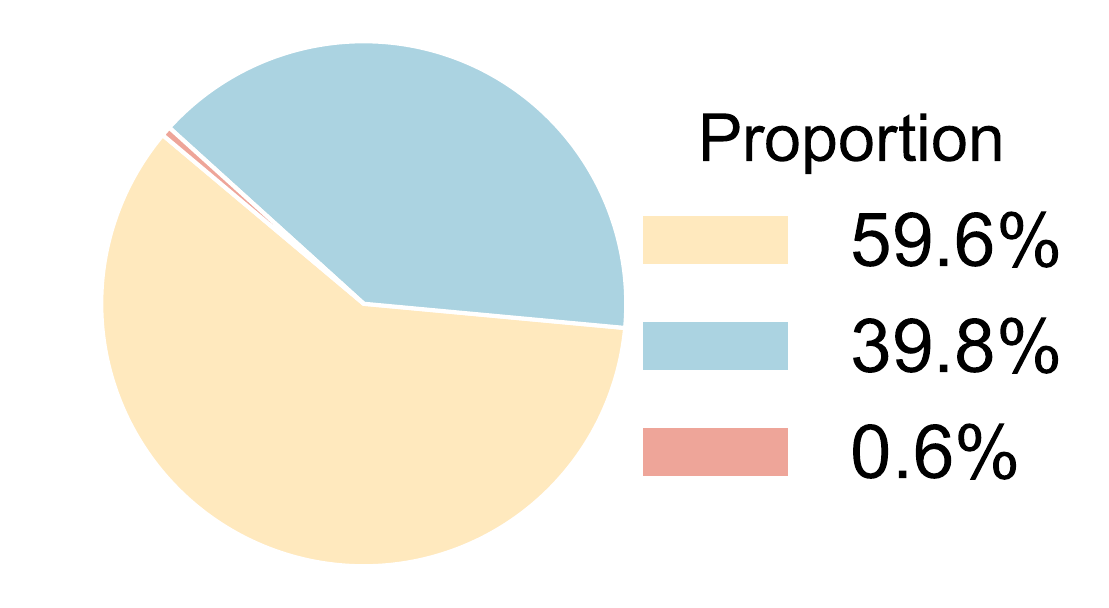}
	}
	
	\caption{Breakdown of Host-to-Device (HtoD) data transfer overhead (Part 2).}
	\label{fig:exp-trans-overhead-part2}
\end{figure}

Figure~\ref{fig:exp-trans-overhead-part2} complements the main text by providing a detailed breakdown of Host-to-Device (HtoD) data transfer overhead for GGNN~\cite{2022GGNN}, GANNS~\cite{2022GANNS}, and SONG~\cite{2020SONG}. These charts further illustrate the substantial proportion of bandwidth consumed by graph topology and raw vector transmission, reinforcing the observation that data movement remains a critical bottleneck for GPU-accelerated ANNS.

\subsection{Supplementary of Search without Transfer}\label{ssec:supplementary-of-search-without-trans}

Figure~\ref{fig:exp-recall-qps-part2} extends the search performance evaluation to a broader range of datasets, plotting QPS against Recall@10 in the high-precision regime. These additional results reinforce the scalability trends observed in the main text, demonstrating how algorithms like CAGRA~\cite{2024CAGRA} sustain superior throughput across diverse data scales and dimensions compared to other GPU baselines.

\subsection{Supplementary of Sub-component in Search}\label{ssec:supplementary-of-subcomp-in-search}

\begin{figure*}[t]
	\vspace{-10pt}
	\includegraphics[width=0.7\textwidth]{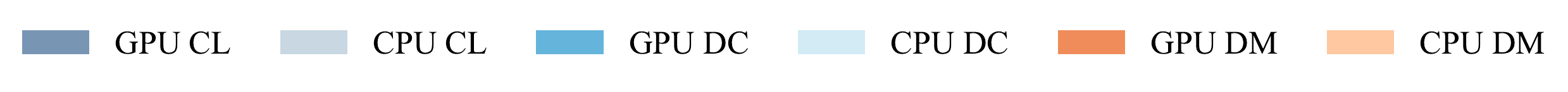}
	\vspace{-10pt}
\end{figure*}

\begin{figure*}[t]
	\vspace{-10pt}
	\centering
	\subfloat[DEEP10M]{
		\includegraphics[width=0.48\linewidth]{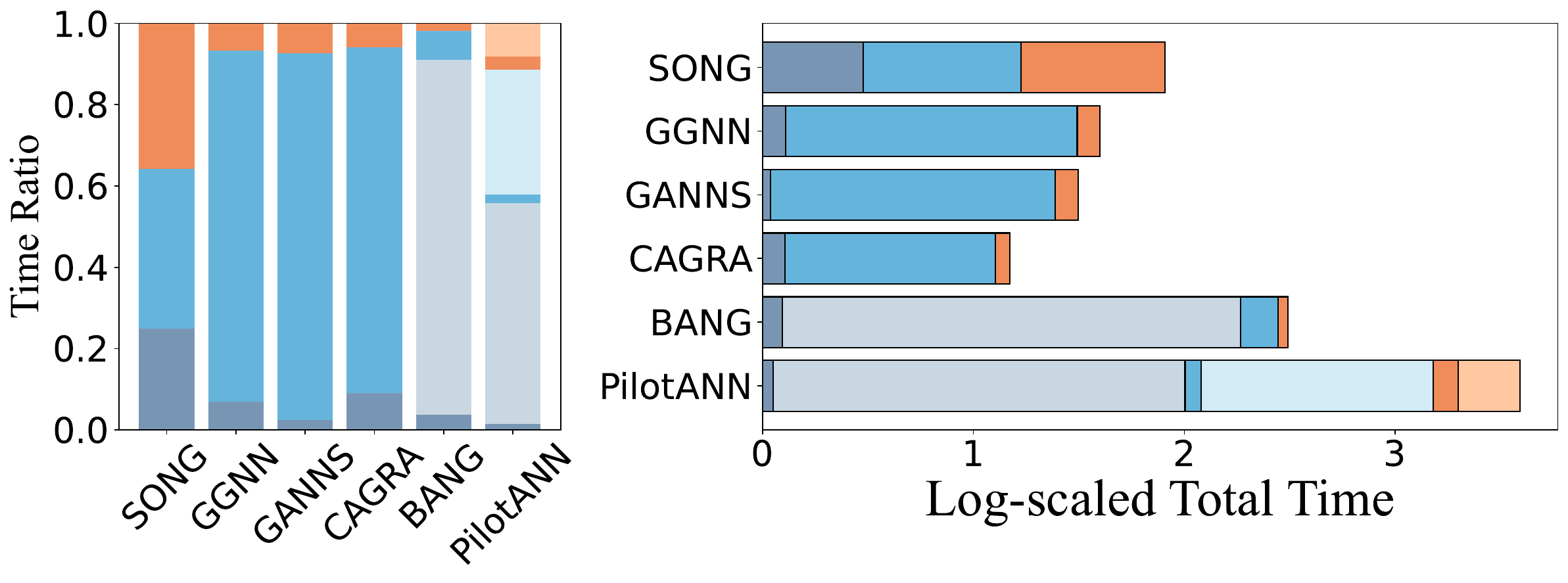}
	}
	\hfill
	\subfloat[SIFT1M]{
		\includegraphics[width=0.48\linewidth]{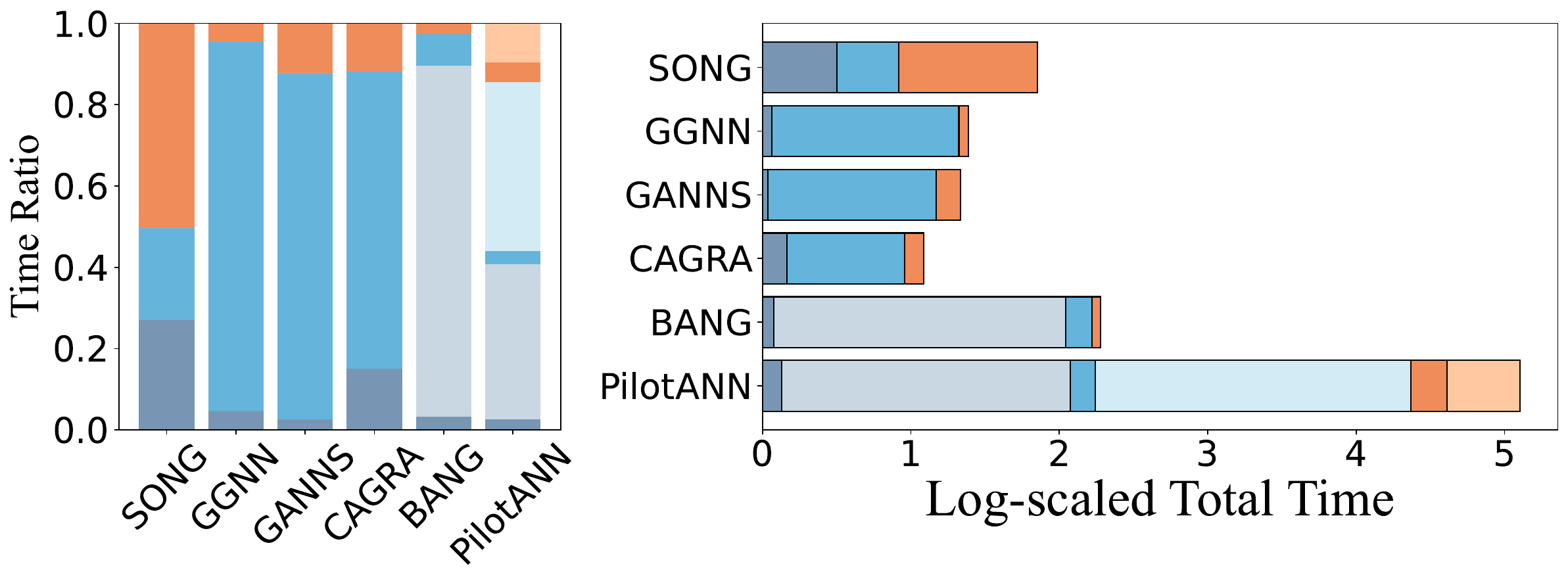}
	}
	\caption{Search component time analysis across different datasets. The charts show the proportion of total kernel time spent on Candidates Locating in Parallel (CL), Bulk Distance Calculation (DC), and Data Structures Maintenance (DM) for various graph-based ANNS algorithms (Part 2).}
	\label{fig:exp-search-components-part2}
\end{figure*}

Figure~\ref{fig:exp-search-components-part2} extends the component-level timing analysis to the DEEP10M and SIFT1M datasets. Consistent with the primary findings, the breakdown reveals that Bulk Distance Calculation (DC) consumes the majority of kernel execution time, consistently overshadowing the costs associated with Candidates Locating (CL) and Data Structures Maintenance (DM).

\end{document}